\documentclass[12pt,reqno]{amsart}

\usepackage{amsmath}

\newtheorem{thm}{Theorem}[section]
\newtheorem{prop}[thm]{Proposition}
\newtheorem{cor}[thm]{Corollary}
\newtheorem{lemma}[thm]{Lemma}

\makeatletter \@addtoreset{equation}{section} \makeatother

\newcommand{\W}{\mbox{${\mathfrak W}$}}
\newcommand{\wat}[1]{\mbox{$\mathfrak{W}(\mathfrak{A}^{#1})$}}
\newcommand{\watt}[1]{{\mbox{\scriptsize ${\mathfrak W}({\mathfrak A}^{#1})$}}}

\newcommand{\ehat}{\mbox{$\widehat{\eta}^{t}_{s}$}}
\newcommand{\enohat}{\mbox{$\eta^{t}_{s}$}}
\newcommand{\winf}{\mbox{${\mathfrak W}^{\infty}$}}
\newcommand{\states}{\mbox{${\mathcal K}\winf$\hspace{1mm}}}

\newcommand{\join}{\mbox{$\bigvee_1 \wat{t}$}}
\newcommand{\dirsys}{\mbox{$\{\wat{t}, \ehat, {\mathbf J}\}$ }}
\newcommand{\threads}{\mbox{$E_{\infty}$}}

\newcommand{\net}{\mbox{$(\mu_t)_{t \in {\mathbf J}}$}\hspace{1mm}}
\newcommand{\C}{\mbox{${\mathcal C}(X)$}}
\newcommand{\Co}{\mbox{${\mathcal C}(X_K)$}}

\newcommand{\R}{\mbox{$\mathbf{\sf R}$}}
\newcommand{\zemu}{\mbox{$\zeta_{\mu}$}}

\newcommand{\g}{\mbox{$\mathfrak{g}$}}
\newcommand{\chip}[1]{\mbox{$\chi^{(X)}_{#1}$}}

\newcommand{\ban}{\mbox{{\bf B}an$_{\mathbf 1}$}}

\newcommand{\Ck}{\mbox{${\mathcal K}C(X_K)$}}
\newcommand{\wo}{\mbox{${\mathfrak W}_K$}}

\newcommand{\spg}{\vspace{5mm} \noindent}
\newcommand{\hk}{\mbox{Haag-Kastler }}
\newcommand{\prf}{\mbox{{\em Proof.\hspace{2mm}}}}
\newcommand{\qd}{\mbox{\hspace{5mm}\rule{2.4mm}{2.4mm}}}
\newcommand{\cc}{compact convex }

\begin{document}

\begin{center}
{\huge {\bf An algebraic theory of infinite classical lattices I:

\vspace{4mm} General theory}}

\vspace{20mm} {\bf Don Ridgeway}

\vspace{15mm}
Department of Statistics,\\
North Carolina State University,\\
Raleigh, NC  27695\\
ridgeway@stat.ncsu.edu
\end{center}

\vspace{1cm} \hspace{53mm} {\bf Abstract}

\vspace{5mm} {\small We present an algebraic theory of  the states of the infinite
classical lattices. The construction follows the Haag-Kastler axioms from quantum field
theory. By comparison, the  *-algebras of the quantum theory are replaced here with the
Banach lattices ($MI$-spaces) to have real-valued measurements, and the
Gelfand-Naimark-Segal construction with the structure theorem for $MI$-spaces to
represent the Segal algebra as $\C$. The theory represents any compact convex set of
states as a decomposition problem of states on an abstract Segal algebra $\C$, where $X$
is isomorphic with the space of extremal states of the set. Three examples are treated,
the study of groups of symmetries and symmetry breakdown, the Gibbs states, and the set
of all stationary states on the lattice. For relating the theory to standard problems of
statistical mechanics, it is shown that every thermodynamic-limit state is uniquely
identified by expectation values with an algebraic state.}

\vspace{1cm} \noindent MSC 46A13 (primary) 46M40 (secondary)

\newpage
\vspace{5mm} \noindent{\bf {\large I \hspace{4mm} Introduction }} \setcounter{section}{1}
\setcounter{equation}{0} \setcounter{thm}{0}

It is now generally recognized in statistical mechanics that in order to well-define even
such basic thermodynamic concepts  as temperature and phase transition, one must deal
with systems of infinite extent \cite{haag96}. Two  approaches to the study of infinite
systems have emerged since the 1950s, Segal's algebraic approach in  quantum field theory
(QFT) (\cite{brat87}, \cite{emch72},  \cite{haag64}, \cite{sega47}) and the theory of
thermodynamic-limit (TL) states (\cite{dobr68},\cite{lanf69},\cite{lanf73}). This paper
is the first of two papers giving an  algebraic theory of measurements on infinite
classical lattices.  In this paper, Part I, we give the general theory. Part II will give
the axiomatic theory of classical measurements. Construction here will be based on a
nonrelativistic variation of the \hk axioms from QFT \cite {haag64}.

Regarding this construction, the observables of an algebraic theory are the elements of a
space satisfying the axioms of the Segal algebra. Example 2 in Segal's original paper
\cite{sega47} is a discussion of the commutative  algebras, the setting for the classical
theory. It shows in particular (Theorem 1) that any commutative  algebra satisfying the
Segal axioms is representable as the space $C(X)$ of real-valued continuous functions on
a certain compact space $X$.  By comparison, in the quantum theory, the observable space
is a $C$*-algebra, and one uses the Gelfand-Naimark-Segal (GNS) construction to represent
it in a standard form as the bounded operators on a certain Hilbert space.

Our space of observables here is a real $MI$-space (Banach lattice with order unit). The
structure theorem for these spaces then provides the representation as a space ${\mathcal
C}(X)$.
 We shall find that the theory focusses on the class of compact convex sets of states on
the Segal algebra. In terms of general statistical mechanics, this important class
includes the set of Gibbs states and the compact sets of states invariant under a group
of symmetries. It also includes the set of all stationary states.

Some of the conclusions about the structure here  are new results of general interest in
statistical mechanics. In particular, we give the proof that the unique Choquet
decomposition of states into extremal states is a general property of the state space of
any infinite lattice. We show, in fact, that any compact convex set of states may be
decomposed into {\em its} extremal states. Although much success has been had in the TL
program in obtaining the decomposability of states in large classes of lattices, the
general proof of this very basic result has not been found. We shall also show that any
TL state is uniquely identifiable by expectation values with an algebraic state.  This
means that the two theories should be regarded as different approaches to a single theory
rather than as different theories.

The material in this paper is arranged as follows. Section II gives the structure of the
lattices themselves and defines the spaces of local observables. Section III introduces
the theory's axioms and applies them to obtain the algebraic observables.  The
representation of algebraic states as threads of local states is the object of Section
IV. It is shown here that this representation enables the identification of TL states
with algebraic states. In Section V, we present the theory of symmetries and symmetry
breakdown, a discussion of Gibbs states. and the construction of the Segal algebra $\C$
for the stationary states of lattices.

\vspace{5mm} \noindent{\bf {\large II \hspace{4mm} The lattice setting  }}
\setcounter{section}{2} \setcounter{equation}{0} \setcounter{thm}{0}

\vspace{3mm}  The purpose of the \hk axioms is to construct an algebraic theory as a
representation of some underlying notion of local observables defined to describe
measurements on a finite (laboratory-scale) system.  Central to the axioms is the {\em
texture} of the theory, in the classical case the assignment, to each such system, of a
space of phase functions representing measurements {\em on that system}. The axioms
define construction of the theory's Segal algebra from its texture. In this section, we
describe the local structure of the lattice in sufficient detail to  define a texture
for it.

\vspace{3mm}
\begin{center} {\bf A. The lattice.} \end{center}

Take the simple lattice gas first.   Denote the infinite lattice by $ \Gamma $,
representable by $ \Gamma = {\mathbb Z}^d$, where $d$ is the dimension of the lattice,
and let $\mathbf T$ be a fixed index set for the lattice sites in $ \Gamma$. Let
$\mathcal P$ denote the set of all finite systems (= bounded subvolumes) of $\Gamma$, and
$\mathbf J$ be a fixed index set for $\mathcal P$. $\mathbf J$ is partially ordered by
set inclusion, {\em i.e.,} for all $s,t \in {\mathbf J}$, write $s \leq t$ iff $\Lambda_s
\subseteq \Lambda_t$, and upward directed by unions.

At any instant, each site is either empty or else it contains a particle. Denote by
$\Omega_o = \{0,1\}$ this set of single-site configurations. For all $i \in {\mathbf T}$,
set $\Omega_i \equiv \Omega_o$ and let $\Omega = {\mathbf{\mathsf P}}_{i \in {\mathbf T}}
\Omega_i$. Then this Cartesian product $\Omega$ is the classical phase space for the
problem. That is, any point $x = (x_i)_{i \in {\mathbf T}} \in \Omega$ gives an
instantaneous configuration of the whole space $\Gamma$.

Now generalize the setting. Hereafter we shall assume only that we have an abstract
infinite lattice system $\Gamma$ and its phase space $\Omega$, together with the
analogous set of bounded systems $\mathcal P$, and its index set {\bf J}.   The more
complicated lattices present nothing new in these terms, although we assume throughout
that the number of single-site configurations is finite. Regarding restrictions that make
certain configurations impossible, these can be introduced either in $\Omega$ itself or
in the distributions assigned in the theory {\em via} the Hamiltonian. The definitions
here require the latter choice.

\vspace{3mm} \begin{center} {\bf B. Local observables.} \end{center}

For the description of a measurement here, we will treat the infinite lattice as
consisting of a system and its infinite surround---a generalized ``temperature
bath''---taking as possible systems of measurement the finite subvolumes of the lattice.
{\em We define the lattice texture  to assure that the expectation values of measurements
on a system are determined by the state of its surround.} This requirement embodies one
of the most basic facts of actual measurements.  In statistical thermodynamics, the
values used in the Gibbs ensembles for the intensive variables of exchange are their
values {\em in the surround}. Thus, for systems that can exchange only heat, Guggenheim
writes, ``$\beta$ $[=1/kT]$ is determined entirely by the temperature bath {\em and so
may be regarded as a temperature scale}'' (\cite[p.65]{gugg57}). A same rule obtains for
the pressure and other variables, and for the same reasons.

To satisfy the requirement, we must be able to define probability distributions (=states)
on the lattice configuration which limit only configurations of the surround, just as we
may freely set the thermostat on a temperature bath, or pressure on a piston. Let
$\mathcal B$ be the Borel $\sigma$-algebra on $\Omega$, and for each finite local system
$\Lambda_t$, let ${\mathfrak A}^t$ be the sub-$\sigma$-algebra generated by sets of the
form $\Omega_{\Lambda_t} \times A_{t^{\prime}}$, where $A_{t^{\prime}} \subseteq
\Omega_{\Lambda_t^{\prime}}$ is any Borel subset of $\Omega_{\Lambda_t^{\prime}}$. Then
for any system $\Lambda_t$, distributions on the probability space $(\Omega, {\mathfrak
A}^t)$ are of the required form.

In order to have expectation values with respect to these states, the local observables
must be  ${\mathfrak A}^t$-measurable. These are functions with preimages satisfying $[f
< a] \equiv \{x \in \Omega: f^t(x) < a\} \in {\mathfrak A}^t \hspace{1mm} \forall a \in
\R$, {\em i.e.,} functions on $\Omega$ with values depending only on configuration {\em
outside} the system $\Lambda^t$. We shall  take as the local observables assigned by the
texture the sets of all bounded Borel-measurable functions on the configuration space of
the lattice subject to this requirement. Because of   prominence of this class of
functions in the theory of Gibbs states in CSM, there is a substantial literature on them
(\cite{ledr73}, \cite{pres76}, \cite{ruel78}).  We adapt the term ``functions {\em from
the outside''} from Preston's monograph \cite{pres76} to describe them.

We define the local observable spaces of the theory as follows.  For all $t \in \mathbf
{J}$, let ${\mathfrak W}_{\circ}({\mathfrak A}^{t})$ be the set of all real-valued,
${\mathfrak A}^{t}$-measurable simple functions on $\Omega$, and denote by ${\mathfrak
W}({\mathfrak A}^{t})$ the uniform closure of ${\mathfrak W}_{\circ}({\mathfrak A}^{t})$
in $l_{\infty}(\Omega)$. Recall that $l_{\infty}(\Omega)$ denotes the Banach space of all
bounded functions on $\Omega$, with sup norm. As constructed, ${\mathfrak W}({\mathfrak
A}^t)$ is the smallest closed linear sublattice in $l_{\infty}(\Omega)$ containing all
characteristic functions. Note in particular that any bounded measurable function can be
obtained as the uniform limit of a sequence of simple functions \cite[page 108]{loev63}.
The Banach space ${\mathfrak W}({\mathfrak A}^{t})$ with its constant function
$\chi_{\Omega}$ defined by $\chi_{\Omega}(x) = 1 \forall x \in \Omega$   is an
$MI$-space, {\em i.e.,} a Banach lattice with order unit. Notationwise, we shall write
elements as $f^t \in \mathfrak{W}({\mathfrak A}^t)$. Throughout the theory, we regard the
elements of $\wat{t}$ as representing measurements on the finite system $\Lambda_t$.

There has long been available an algebraic theory for classical infinite systems
\cite{ruel69}. In contrast with the preceding, the probability algebras are defined
analogously so that local observables in this theory are functions on $\Omega$ have
values depending only on configurations {\em inside} $\Lambda_t$. As we shall see, this
leads to a much simpler construction of the Segal algebra. Obviously, our lattice
requirement would not be satisfied in these algebras.

\vspace{3mm} \begin{center} {\bf C. Local states.} \end{center}

We complete our discussion of the local structure by introducing the local state spaces.
Denote the set of states on $\wat{t}$ by $E_t$, {\em i.e.,} the set of all positive
linear functionals on $\mathfrak{W}({\mathfrak A}^t)$ with norm 1. Notationwise, write
$\mu_t \in E_t$. In terms of the theory of measurement, we choose a particular system
$\Lambda_t$ and fix the state of the lattice $\mu_t \in E_t$.  By construction, this
choice affects only configurations in the surround. Then the expectation value of any
measurement $f^t \in \wat{t}$ is $\mu_t(f^t)$. This is the formal equivalent of
determining the value of a measurement by setting the temperature of the heat bath as a
reading on its thermostat.

The characterization of states is contained in the following proposition. For this
result, denote by $\bigcirc$*\wat{t} the unit ball of the (topological) dual of \wat{t}
with its wk*-topology. The notation is from category theory. Unless otherwise stated, the
topology on $E_t$ in the following is the wk*-topology.  By compact, we shall always mean
wk*-compact Hausdorff.

\begin{prop}
The linear functional  $\mu_{t}$ on $\wat{t}$ is a state iff $\|\mu_{t}\| =
\mu_{t}(\chi_{\Omega}) = 1$. The set of all states on $\wat{t}$, denoted $E_{t}$, is a
nonempty compact subset of the unit ball $\bigcirc$*$\wat{t}$.
\end{prop}

\noindent \prf   Suppose $\mu_t$ is a state on $\wat{t}$.   We must show
$\mu_t(\chi_{\Omega}) = 1.$ For all $f^t \in \wat{t}$such that $0 \leq f^t \leq 1,$
$\mu_t(\chi_{\Omega} - f^t) = \mu_t(\chi_{\Omega}) - \mu_t(f^t) \geq 0$, and therefore $1
= \|\mu_t \| \equiv \sup_{\| f^t \| \leq 1}$ $ |\mu_t(f^t)| = \mu_t (\chi_{\Omega})$.
Conversely, suppose $\mu_t$ is a linear functional on $\wat{t}$ such that $\|\mu_t\| =
\mu_t(\chi_{\Omega}) = 1.$ We must show that $\mu_t \geq 0$.  Fix any $f^t$ such that $0
\leq f^t \leq \chi_{\Omega}.$  Then $0 \leq \chi_{\Omega} - f^t \leq 1$ and therefore
$\|\chi_{\Omega} - f^t \| \leq 1.$  Then $|\mu_t(\chi_{\Omega} - f^t)| =
|\mu_t(\chi_{\Omega}) - \mu_t(f^t)| = |1 - \mu_t(f^t)| \leq 1$ because $\mu_t$ is
contracting, and the conclusion follows.

The set $E_{t}$ is always nonempty  $\hspace{.5em}\forall t \in \mathbf{J}.$ In fact, for
any $x \in \Omega$, define the {\bf point functional} $\delta_{x}: \wat{t} \rightarrow$
{\bf {\sf R}} by $\delta_{x}f^{t} = f^{t}(x).$ Then clearly, $\delta_{x}$ is a positive
linear functional, and $\| \delta_{x} \| = \delta_{x}(\chi_{\Omega}) =
1\hspace{.5em}\forall x \in \Omega.$ Hence, $\delta_{x} \in E_{t}\hspace{.5em}\forall x
\in \Omega.$

Now define the linear function ${\mathfrak z} : (\wat{t})$*$ \rightarrow$ {\bf {\sf R}}
by ${\mathfrak z}(\mu) = \mu(\chi_{\Omega})$. Then ${\mathfrak z}$ is wk*-continuous.  In
fact, for any $\varepsilon >0$, let $U = {\mathcal N}(\mu; \chi_{\Omega}, \varepsilon)$
be the subbasic open set $\{\nu \in (\wat{t})$*$ : |\nu(\chi_{\Omega}) -
\mu(\chi_{\Omega})|<\varepsilon\}.$ Then ${\mathcal N}(\mu; \chi_{\Omega}, \varepsilon) =
\{ \nu : {\mathfrak z}(\nu) \in ({\mathfrak z}(\mu) - \varepsilon, {\mathfrak z}(\mu) +
\varepsilon)\}$.  Hence, for any $\mu \in (\wat{t})$*$, \varepsilon > 0, {\mathfrak
z}({\mathcal N}(\mu; \chi_{\Omega}, \varepsilon)) \subseteq ({\mathfrak z}(\mu) -
\varepsilon, {\mathfrak z}(\mu) + \varepsilon)$. Then ${\mathfrak z}^{\leftarrow}(1)$ is
closed. The unit ball $\bigcirc$*$\wat{t}$ is wk*-compact by Alaoglu's Theorem.  Since
$E_{t}$ is the intersection ${\mathfrak z}^{\leftarrow}(1) \cap \bigcirc$*$\wat{t}$ of a
closed hyperplane and a compact set, it is closed and compact. \qd

\setcounter{section}{3} \vspace{5mm} \noindent{\bf {\large III \hspace{4mm} Algebraic
theory }} \setcounter{section}{3} \setcounter{equation}{0} \setcounter{thm}{0}

\vspace{3mm} \begin{center} {\bf A. The Haag-Kastler frame.} \end{center}

{\bf 1. The axioms.} We are now in a position to formulate the Haag-Kastler frame for the
classical case. For nonrelativistic theory, it is defined by four axioms. We state the
axioms in terms of the above structure.

\spg {\bf Axiom 1} The lattice texture is defined by the pairings

\begin{equation}
\Lambda_t \mapsto \wat{t},\hspace{5mm} t \in {\mathbf J}
\end{equation}

\spg {\bf Axiom 2} Define the order relation $\leq$ on the net $(\wat{t})_{t \in {\mathbf
J}}$ by $\wat{s} \leq \wat{t}$ iff $\wat{s} \supseteq \wat{t}$.  Then the net
$(\wat{t})_{t \in {\mathbf J}}$ is an upward-directed partially ordered set, with $s \leq
t \Rightarrow \wat{s} \leq \wat{t}$.

\spg {\bf Axiom 3} All local observables are compatible.

\spg {\bf Axiom 4} The theory's Segal algebra is the completion of the inductive limit of
the net of local algebras $(\wat{t})_{t \in {\mathbf J}}$. It is representable as a space
$\C$ for a suitable compact space $X$.

\spg Axiom 1 identifies each system $\Lambda_t$ with the corresponding $MI$-space of
observables from outside $\Lambda_t$. Note especially that the local algebras $\wat{t}$
are defined without reference to (or need for) containing walls for the systems
$\Lambda_t$.  Axiom 2 is an order structure imposed by the texture (3.1). Order by
inclusion among the systems $\Lambda_t$ defines a partial order of the local observable
algebras as well by (1.1). It follows from the definition of the ${\mathfrak A}^{t}$ that
for all $\Lambda_s \subseteq \Lambda_t$, $\wat{s} \supseteq \wat{t}$. Axiom 3 has to do
with the compatibility of observables on different systems. This is classical theory. The
final axiom constructs the theory's Segal algebra, the space of {\em quasilocal
observables}, as the completion of an inductive limit. We shall prove that the completion
of this limit is an $MI$-space. This will assure its representation as $\C$
\cite[13.1.1.]{sema71}.

\vspace{3mm} {\bf 2. The morphisms} $(\ehat)_{s \leq t}$. The inductive limit in Axiom 4
will be in the category {\bf Ban}$_1$ of Banach spaces and linear contractions. The
definition of the limit requires the upward-directed net of $MI$-spaces and, for each
nested pair of systems $\Lambda_s \subset \Lambda_t$, a morphism mapping measurements on
the smaller system to those measuring the same physical quantity on the larger system,
i.e., a set of positive linear contractions  with the following properties:

{\renewcommand{\labelenumi}{(\roman{enumi})}
\begin{enumerate}
\item $\forall f^{t} \in {\mathfrak W}({\mathfrak A}^{s}) \cap \wat{t}$,
$\widehat{\eta}^{t}_{s}(f^{t}) = f^{t}$;

\item $\forall t \in {\bf J}$, $\widehat{\eta}^t_t f^t = f^t$; and

\item the composition rule: $\forall r \leq s \leq t$, $\widehat{\eta}^t_s
\widehat{\eta}^s_r = \widehat{\eta}^t_r.$
\end{enumerate}
}

\noindent Since the form of  these functions is unknown,  the axiom requires their
existence by assumption. Certainly thermodynamics requires that it be possible to
identify observables on different systems that measure the same physical quantity.

To gain familiarity with the morphisms, it is useful to study their properties with a set
of functions that approximate their action. For all $s,t \in {\bf J}$, we define $\ehat:
\wat{s} \rightarrow \wat{t}$ as follows. For all $s \in {\bf J}$, $\widehat{\eta}^s_s$ is
the identity mapping. For $s < t$, denote by $M^t_s = \Omega_{\Lambda_t \sim \Lambda_s}$,
and let

\begin{equation} \ehat f^s(x) = \frac{1}{|M^t_s|}\sum_{(M^t_s)} f^s(x), \hspace{1cm} x \in
\Omega, f^s \in \wat{s}
\end{equation}

\noindent The sum notation $(M^t_s)$ means to sum over all configurations in $M^t_s$,
holding the rest of $x \in \Omega$ constant, and $|M^t_s|$ is the number of such
configurations.  The effect of the mappings $\ehat$ is to remove the dependence on
configurations in $M^t_s$ by averaging over them. As for the approximation, note in
particular that $f^s \in \wat{s}$ and $\ehat f^s$ differ as functions on the infinite
space $\Omega$ at most at configurations in the finite region  $\Lambda_s \sim
\Lambda_t$, where the average is performed. The first two properties are immediate. For
the composition rule, one has the following:

\begin{equation}
\begin{array}{cl}

\widehat{\eta}^t_rf^r(x) & = \frac{1}{|M^t_r|} \sum_{(M^t_r)} f^r(x)\\
& = \frac{1}{|M^t_s|\cdot|M^s_r|}\sum_{(M^t_s)} \sum_{(M^s_r)} f^r(x) = \frac{1}{|M^t_s|}
\sum_{(M^t_s)}
\frac{1}{|M^s_r|}\sum_{(M^s_r)}f^r(x)\\
& = \frac{1}{|M^t_s|} \sum_{(M^t_s)} \widehat{\eta}^s_rf^r(x) =
\widehat{\eta}^t_s\widehat{\eta}^t_rf^r(x).

\end{array}
\end{equation}

\noindent The similarity in properties of the function $\ehat f^s$ to a conditional
expectation is apparent, although here the smoothing action is independent of state.

From the $(\ehat)_{s \leq t}$ we may immediately construct a corresponding set of
morphisms relating local state spaces. This is obtained  as follows.

\begin{prop}   For each comparable pair $s \leq t$, define the
wk*-continuous mapping $\enohat: \bigcirc$*$\wat{t}
 \rightarrow \bigcirc$*$\wat{s}$ by $\enohat \mu_{t}
= \mu_{t} \circ \widehat{\eta}^{t}_{s}$.  The mapping \enohat
carries states to states and non-states to non-states. Moreover,
$\eta^{t}_{s}$ is a wk*-continuous mapping on $E_{t}$ into
$E_{s}$.  For all $r \leq s \leq t, \eta^{s}_{r} \eta^{t}_{s} =
\eta^{t}_{r}$, and $\eta^{t}_{t}$ is the identity mapping
$\iota_{E_{t}}$.
\end{prop}

\noindent \prf Fix any state $\mu_{t} \in E_{t}$. By hypothesis, $\widehat{\eta}^{t}_{s}
\chi_{\Omega} = \chi_{\Omega}$, so that $\enohat \mu_{t}(\chi_{\Omega}) =
\mu_{t}(\widehat{\eta}^{t}_{s} \chi_{\Omega}) = \mu_{t}(\chi_{\Omega}) = 1$.  Moreover,
$\forall\hspace{.3em}\mu_{t} \in E_{t}, \|\enohat \mu_{t} \| = \sup_{\|f^{s}\| \leq 1}$
$|\enohat \mu_{t} (f^{s})|  = \sup_{\|f^{s}\| \leq 1} |\mu_{t}(\widehat{\eta}^{t}_{s}
f^{s})|$. Since $|\widehat{\eta}^{t}_{s}f^{s}| \leq |f^{s}|$, $\sup_{\|f^{s}\| \leq 1}
|\mu_{t}(\widehat{\eta}^{t}_{s} f^{s})| \leq \sup_{\|f^{t}\| \leq 1}|\mu_{t}(f^{t})| =
\|\mu_{t}\| = 1$. But $\sup_{\|f^{s}\| \leq 1}$ $ |\enohat \mu_{t} (f^{s})| \geq \enohat
\mu_{t} (\chi_{\Omega}) = \mu_{t}(\chi_{\Omega}) = 1$. Then $\|\enohat \mu_{t} \| = 1$.
Hence, $\enohat \mu_{t}$ is a state by Proposition II.1.  Now suppose $\mu_{t} \in
\bigcirc$*${\mathfrak W}({\mathfrak A}^{t})\backslash E_{t}$.  If $\mu_t$ is not a
positive functional, then there exists $f^t \in \wat{t}$, $f^t \geq 0$, such that
$\mu_t(f^t) < 0.$ Then $\enohat \mu_t(f^t) = \mu_t(\ehat f^t) = \mu_t(f^t) < 0,$ so that
$\enohat \mu_t$ is not a positive functional.  If $\mu_{t}(\chi_{\Omega}) < 1$, $\enohat
\mu_{t}(\chi_{\Omega}) = \mu_{t}(\chi_{\Omega}) < 1.$ In either case, $\enohat \mu_t \not
\in E_s$. Finally, $\forall f^t \in \wat{t}$, $\mu_t(f^t) = \mu_t(\widehat{\eta}^t_t f^t)
= \eta^t_t \mu_t(f^t).$  The wk*-continuity is shown in \cite[Proposition 6.1.8]{sema71}.
\qd

We have defined two systems, $\{\wat{t}, \ehat, {\bf J} \}$ and $\{E_t, \enohat,$
$\mathbf{J} \}$. The set $\{\wat{t}, \ehat, {\bf J} \}$ is an inductive system in the
category $\ban$. The set $\{E_t, \enohat, \mathbf{J} \}$ is a projective system in the
category {\bf C}ompconv of compact convex spaces and continuous affine maps. The notation
follows Semadeni \cite{sema71}.

\vspace{3mm} {\bf 3. Notation.} As regards notation, the transformations
$\widehat{\eta}^{t}_{s}: {\mathfrak W}({\mathfrak A}^{s}) \rightarrow {\mathfrak
W}({\mathfrak A}^{t})$ require two indices, specifying domain and range.  The form of the
superscript/subscript notation follows the conventions of tensor contraction.  Thus, in
$\widehat{\eta}^{t}_{s}f^{s}$, the index $s$ cross-cancels to take a function $f^{s} \in
\wat{s}$ with superscript $s$ over to a function in ${\mathfrak W}({\mathfrak A}^{t})$ with
superscript $t$. Similarly, one encounters later $\eta^{t}_{s}\mu_{t}$, in which the $t$
cross-cancels to take the state $\mu_{t}$ with subscript $t$ to a new state with subscript
$s$. In the compose $\widehat{\eta}^{t}_{s} \widehat{\eta}^{s}_{r}$, the $s$ cross-cancels
to leave a transformation $\widehat{\eta}^t_r$ with superscript $t$, subscript $r$.  Also
encountered will be $\mu_t = \rho_{t} \mu$, taking an object $\mu$ with no index to one
with subscript $t$, as well as $\sigma_{s}f^{s} = [f]$, taking an indexed function $f^t$ to
$[f]$ with no index.

\begin{center} {\bf B. The inductive limit $\lim^{\rightarrow}\{\wat{t}, \ehat, {\mathbf
J}\}$.}
\end{center}

We begin with the first part of Axiom 4, the construction of the inductive limit from the
lattice texture.

\begin{thm}
The system $\{\wat{t}, \ehat, {\mathbf J}\}$ has a unique inductive limit $\{\winf,$ $
\sigma_t, {\mathbf J}\} = \lim^{\rightarrow}\{\wat{t}, \ehat, {\mathbf J}\}$.  \winf is a
Banach space, and the $\sigma_{t}:{\mathfrak W}({\mathfrak A}^{t}) \rightarrow {\mathfrak
W}^{\infty}$ are linear contractions obeying the composition rule $\sigma_{s} = \sigma_{t}
\widehat{\eta}^{t}_{s}\hspace{.5em}\forall s \leq t$.
\end{thm}

\noindent \prf The properties  of the morphisms $(\ehat)$   assure that the set
$\{\wat{t}, \ehat, t \geq s, s \in \mathbf{J} \}$ is a inductive system of Banach spaces,
characterized by the commuting diagram:

\setlength{\unitlength}{.25mm}
\begin{picture}(250,200)(-140,0)
\put(25,90){${\mathfrak W}({\mathfrak A}^{t})$}
\put(140,20){${\mathfrak W}({\mathfrak A}^{r})$}
\put(140,165){${\mathfrak W}({\mathfrak A}^{s})$}
\put(85,40){$\widehat{\eta}^{t}_{r}$}
\put(160,95){$\widehat{\eta}^{s}_{r}$}
\put(85,145){$\widehat{\eta}^{t}_{s}$}
\put(140,25){\vector(-4,3){80}} \put(140,165){\vector(-4,-3){80}}
\put(155,35){\vector(0,1){120}}
\end{picture}

\noindent for all $r \leq s \leq t$.  A standard construction of
the Banach-space limit applies to the system $\{ {\mathfrak
W}({\mathfrak A}^{t}), \widehat{\eta}^{t}_{s}, \mathbf {J} \}$ as
follows\cite[p. 212]{sema71}. Let $\bigvee_{1} {\mathfrak
W}({\mathfrak A}^{t})$ be the $l_{1}$-join of the individual
algebras, {\em i.e.}, the Banach space $\{ (f^{t})_{t \in \mathbf{
J}} \in ${\bf {\sf P}}${}_ {t \in \mathbf{J}}{\mathfrak
W}({\mathfrak A}^{t}): \sum \|f^{t} \|_{t} < \infty \}$, with the
usual linear operations. The norm is $\| f \| = \sum \|f^t
\|_{t}$, where $\| f^{t}\|_{t}$ is the norm on ${\mathfrak
W}({\mathfrak A}^{t})$.  The $l_{1}$-join is ordered by the
relation $(f^{t}) \leq (g^{t})$ iff $f^{t} \leq
g^{t}$\hspace{.5em} $\forall t \in \mathbf{ J}$. Let
$\widehat{\sigma}_{t}: {\mathfrak W}({\mathfrak A}^{t})
\rightarrow \bigvee_{1}{\mathfrak W}({\mathfrak A}^{t})$ be the
canonical injection, and ${\mathfrak M}$ be the closed linear
subspace of $\bigvee_{1}{\mathfrak W}({\mathfrak A}^{t})$ spanned
by elements of the form $\widehat{\sigma}_{s}(f^{s}) -
\widehat{\sigma}_{t}(\widehat{\eta}^{t}_{s}f^{s})$, $f^{s}  \in
{\mathfrak W}({\mathfrak A}^{s}), s \leq t$.  Then for the
inductive limit $\{ {\mathfrak W}^{\infty}, \sigma_{t}, \mathbf{
J}\}$, the object ${\mathfrak W}^{\infty}$ is the quotient space
$\bigvee_{1}{\mathfrak W}({\mathfrak A}^{t})/{\mathfrak M}$.
Denote the elements of ${\mathfrak W}^{\infty}$ with square
brackets.  \winf has the usual quotient norm
$\|\hspace{.2em}[f]\hspace{.2em}\| = \inf_{h \in {\mathfrak M}}\|f
+ h \|$. Let $\tau: \bigvee_{1}{\mathfrak W}({\mathfrak A}^{t})
\rightarrow {\mathfrak W}^{\infty}$ be the quotient surjection.
The limit homomorphism $\sigma_{t}:{\mathfrak W}({\mathfrak
A}^{t}) \rightarrow {\mathfrak W}^{\infty}$ is the compose
$\sigma_{t} = \tau \circ \widehat{\sigma}_{t}$. All elements of
${\mathfrak W}^{\infty}$ are of the form $[f] = \sum
\sigma_{t_{k}}f^{t_{k}}$ for some countable set of functions
$f^{t_{k}} \in {\mathfrak W}({\mathfrak A}^{t_{k}})$. The
composition rule for the $(\sigma_t)_{t \in {\mathbf J}}$ makes
the following diagram commuting, for all $s \leq t$:

\setlength{\unitlength}{.25mm}
\begin{picture}(250,200)(-140,0)
\put(40,90){${\mathfrak W}^{\infty}$} \put(140,20){${\mathfrak
W}({\mathfrak A}^{s})$} \put(140,165){${\mathfrak W}({\mathfrak
A}^{t})$} \put(85,40){$\sigma_{s}$}
\put(160,95){$\widehat{\eta}^{t}_{s}$} \put(85,145){$\sigma_{t}$}
\put(140,25){\vector(-4,3){80}} \put(140,165){\vector(-4,-3){80}}
\put(155,35){\vector(0,1){120}}
\end{picture}

\qd

\spg In introducing the morphisms $(\ehat)$, we identified $f^s$ and its image $\ehat
f^s$ as physically equivalent local measurements.  The formation of $\winf$ as the
quotient space assures that equivalent measurements map to the same quasilocal
observable, $\sigma_sf^s = \sigma_t\ehat f^s$. We use the notation $\phi \in \winf$.

\begin{center} {\bf  C. Functional representation of \winf.}
\end{center}

Axiom 4 calls for construction of the Segal algebra as an $MI$-space formed by completion
of this inductive limit. Of course, as a Banach space, $\winf$ is complete with respect
to its norm topology, but it is not an $MI$-space. We shall show that its functional
representation as an order-unit space satisfies this condition as well. We first provide
the three main elements needed for this construction, namely, a partial order in $\winf$,
an order unit in $\winf$, and the set of states ${\mathcal{K}}(\winf)$ on $\winf$.

{\bf 1. Order structure of $\winf$.} We assign the usual {\bf quotient partial order} to
${\mathfrak W}^{\infty}$ determined by the surjection $\tau$. Explicitly, one writes $[g]
\leq [f]$ iff there exists a finite set of pairs $(a_{i},b_{i}) \in \bigvee_{1}
{\mathfrak W}({\mathfrak A}^{t})$ such that (i) $a_{i} \leq b_{i} \hspace{.5em}\forall i
= 1, \dots , n$, (ii) $ [g] =[a_{1}]$ and $[b_{n}] = [f]$, and (iii) $[b_{1}] = [a_{2}],
\dots , [b_{n - 1}] = [a_{n}]$.\cite[Definition 2.3.4] {sema71} For example, with $n =
2$, this leads to $[g] = [a_{1}] \leq [b_{1}]  = [a_{2}] \leq [b_{2}] = [f]$.  Note in
particular that for $n =1$, $[g] \leq [f]$ iff there exists $a \leq b$ such that $[a] =
[g],$ $[b] = [f].$

The induced order relation has the following properties.

\begin{lemma}
$[g] \leq [f]$ iff for any $g \in \tau^{\leftarrow}[g], f \in
\tau^{\leftarrow}[f]$ there exists an $h \in {\mathfrak M}$ such
that $g \leq f+h$.  Moreover, the surjection $\tau$ is
order-preserving, i.e., $\tau(C) = [C]$, where $C$ and $[C]$ are
the positive cones in $\bigvee_{1} {\mathfrak W}({\mathfrak
A}^{t})$ and ${\mathfrak W}^{\infty}$, respectively.
\end{lemma}
\noindent
\prf
Suppose that the inequality $[0] \leq [f]$ is determined
by $n$ pairs $(a_{i},b_{i})$. Define the set of elements $(h_{i})
\in {\mathfrak M}$ by $h_{1} = -a_{1}, h_{n+1} = b_{n}-f$, and
$h_{i} = b _{i - 1} - a_{i}\hspace{.5em}\forall i = 2, \ldots, n$.  Then from $a_{i} \leq b_{i}
\hspace{.5em}\forall  i, \sum_{i=1}^{n} a_{i} \leq \sum_{i=1}^{n} b_{i}$, and thus,
$0 \leq \sum_{i=1}^{n} (b_{i}-a_{i})$.  Hence, $0 \leq f-a_{1} + \sum_{i=2}^{n}
(b_{i-1}-a_{i}) + (b_{n}-f)$, or $0 \leq f + \sum_{i=1}^{n+1} h_{i}$.
Set $h = \sum_{i=1}^{n+1} h_{i} \in {\mathfrak M}$.  For the converse, one
has immediately that $f \geq h \in {\mathfrak M}$ implies (by definition) that
$[f] \geq [h] = [0]$.  Thus, $[0] \leq [f]$ iff there exists $h \in
{\mathfrak M}$ such that $0 \leq f + h$.  Now apply this result.  For any pair $[g], [f]$, $[g] \leq [f]$
or $[0] \leq [f]
- [g] = [f - g]$ iff there exists $h \in {\mathfrak M}$ such that $0 \leq
(f-g) + h$ or $ g \leq f+ h$.  Finally, to show that $\tau(C) \supseteq
[C]$, fix any $[g] \in [C]$.  Then for all $f \in \tau^{\leftarrow}[g],$ there exists $h
\in {\mathfrak M}$ such that $f + h \in C$.  But $\tau(f+h) = [f + h] =
[f] = [g].$  Conversely, $\tau(C) \subseteq [C]$ because  $f \geq 0$
implies that $[f] \geq [0].$  Hence, $\tau (C) = [C].$
\qd

\vspace{3mm} {\bf 2. The order unit $e \in \winf$}. The next result identifies an element
$e \in \winf$ with special properties.

\begin{thm}
Fix any $t \in \mathbf{J}$, and let $ e  = \sigma_{t}(\chi_{\Omega})$.
Then $e$ is independent of
the choice of $t$.  The element $e$ is an order unit for the space \winf, so that
for every $[f] \in {\mathfrak W}^{\infty}, -\|\hspace{.3em}[f]\hspace{.3em}\|e
\leq [f] \leq \|\hspace{.3em}[f]\hspace{.3em}\| e$.
\end{thm}
\noindent \prf Recall first the definition of an order
unit.\cite[p.68ff]{alfs71}  Let $A$ be an ordered linear space.
The linear subspace $J \subseteq A$ is called an {\bf order ideal}
iff for all $a, b \in J$ and $c \in A$, the inequality $a \leq c
\leq b$ implies that $c \in J$, For any $a \in A$, denote by
$J(a)$ the smallest order ideal containing $a$.  Then $a$ is said
to be an {\bf order unit} of A if $J(a) = A$.

We obtain a general form for the elements of ${\mathfrak W}^{\infty}$, namely, that
$\winf =\{\sum^{\infty} \sigma_{t_{k}}(f^{t_{k}}) \in \mbox{{\bf {\sf P}}}_{t \in
\mathbf{J}}\sigma_{t}({\mathfrak W}({\mathfrak A}_{t})): \sum^{\infty} \|f^{t_{k}}
\|_{t_{k}} < \infty\}$.  Indeed, by definition, everything in ${\bf
\bigvee_{1}}{\mathfrak W}({\mathfrak A}^{t})$ is of the form $\sum^{\infty}
\widehat{\sigma}_{t_{k}}(f^{t_{k}})$ for some countable set of indices $(t_{k})$.  The
quotient surjection $\tau : \bigvee_{1}{\mathfrak W}({\mathfrak A}^{t}) \rightarrow
{\mathfrak W}^{\infty}$ is linear and of norm $\|\tau\| = 1$.  From linearity,
$\tau(\sum^{n} \widehat{\sigma}_{t_{k}}f^{t_{k}}) = \sum^{n} \sigma_{t_{k}}f^{t_{k}} \/
\forall n \in$ {\bf {\sf N}}, since $\sigma_{t} = \tau \circ \widehat{\sigma}_{t}$.  Then
from continuity, $\tau \sum^{\infty}\widehat{\sigma}_{t_{k}}(f^{t_{K}}) = \sum^{\infty}
\sigma_{t_{k}}(f^{t_{k}})$.  But $\tau$ is onto.  Hence, everything in ${\mathfrak
W}^{\infty}$ is attained in this way.

Fix any $t \in \mathbf{J}$, and define $e = \sigma_t(\chi_{\Omega})$.
Then for any other $s \in \mathbf{J}$, there exists $u \in
\mathbf{J}$ such that $s,t \leq u$, because {\bf J} is upward
directed. Then $\sigma_t(\chi_{\Omega}) - \sigma_{u}(\chi_{\Omega}) =
\sigma_t(\chi_{\Omega}) - \sigma_{u}(\widehat{\eta}^u_t\chi_{\Omega})
\in \mathfrak{M}$, so that $\sigma_{u}(\chi_{\Omega}) =
\sigma_t(\chi_{\Omega}) = e$ and similarly, $\sigma_t(\chi_{\Omega}) =
e$. Hence, the definition $e = \sigma_t(\chi_{\Omega})$ is
independent of $t$.

We construct a more general element. Fix any $f = \sum_{k = 1}^{n}
\widehat{\sigma}_{t_{k}} (f^{t_{k}}) \in {\bf \bigvee_{1}}
{\mathfrak W}({\mathfrak A}^{t})$, with $n$ finite, and consider
the sum $\sum_{k = 1}^{n} (\|f^{t_{k}}\|_{t_{k}}/\|f\|)
\widehat{\sigma}_{t_{k}} (\chi_{\Omega}))$.  For any $u \geq t_k\
\forall k=1,\ldots,n$,

\[ \begin{array}{l}
\sum_{k = 1}^{n} (\|f^{t_{k}}\|_{t_{k}}/\|f\|)
\widehat{\sigma}_{t_{k}} (\chi_{\Omega})) - \widehat{\sigma}_u(1_u)
=\\ \sum_{k = 1}^{n} (\|f^{t_{k}}\|_{t_{k}}/\|f\|)
\widehat{\sigma}_{t_{k}} (\chi_{\Omega})) - \sum_{k = 1}^{n}
(\|f^{t_{k}}\|_{t_{k}}/\|f\|) \widehat{\sigma}_u (\chi_{\Omega}) =\\
\sum_{k = 1}^{n} (\|f^{t_{k}}\|_{t_{k}}/\|f\|)
(\widehat{\sigma}_{t_{k}} (\chi_{\Omega}) -
\widehat{\sigma}_u(\widehat{\eta}^u_{t_k}\chi_{\Omega})).
\end{array}  \]

\spg But $\widehat{\sigma}_{t_k} (\chi_{\Omega}) -
\widehat{\sigma}_u(\widehat{\eta}^u_{t_k}\chi_{\Omega}) \in \mathfrak{M}$ for all $t \in
\mathbf{J}$. Since ${\mathfrak M}$ is a linear subspace, it contains all linear
combinations of its elements. Hence, $\sum_{k = 1}^{n} (\|f^{t_{k}}\|_{t_{k}}/\|f\|)
\widehat{\sigma}_{t_{k}} (\chi_{\Omega}))$ $ = e$. Then for any countable set of indices
$(t_{k}) \in \mathbf{J}$, the infinite sum converges,

\[ \sum_{k = 1}^{\infty} (\|f^{t_{k}}\|_{t_{k}}/\|f\|)
\widehat{\sigma}_{t_{k}} (\chi_{\Omega})) = e, \]

\spg because the equivalence classes are closed.

Now fix any $f \in \bigvee_{1}{\mathfrak W}({\mathfrak A}^{t})$ and any $h \in {\mathfrak
M}$.  There exists a countable set of indices $(t_{k}) \in \mathbf{J}$ such that $f+h =
\sum_{k=1}^{\infty} \widehat{\sigma}_{t_{k}}(f^{t_{k}} + h^{t_{k}})$. Clearly,  $-a \sum
\|f^{t_{k}} + h^{t_{k}}\|_{t_{k}} \widehat{\sigma}_{t_{k}}(\chi_{\Omega}) \leq \sum
\widehat{\sigma}_{t_{k}}(f^{t_{k}} + h^{t_{k}}) \leq a \sum \|f^{t_{k}} +
h^{t_{k}}\|_{t_{k}} \widehat{\sigma}_{t_{k}}(\chi_{\Omega}) $ iff $a \geq 1$.  Then
$-b(h)\sum (\|f^{t_{k}} + h^{t_{k}}\|_{t_{k}}/\|f+h\|)
\widehat{\sigma}_{t_{k}}(\chi_{\Omega}) \leq \sum \widehat{\sigma}_{t_{k}}(f^{t_{k}} +
h^{t_{k}}) \leq b(h) \sum (\|f^{t_{k}} + h^{t_{k}}\|_{t_{k}}/\|f+h\|)
\widehat{\sigma}_{t_{k}}(\chi_{\Omega}) $ iff $ b(h) \geq \|f+h\|$. Therefore, there exists
$h \in {\mathfrak M}$ such that $-b\sum (\|f^{t_{k}} + h^{t_{k}}\|_{t_{k}}/\|f+h\|)
\widehat{\sigma}_{t_{k}}(\chi_{\Omega}) \leq \sum \widehat{\sigma}_{t_{k}}(f^{t_{k}} +
h^{t_{k}}) \leq b \sum (\|f^{t_{k}} + h^{t_{k}}\|_{t_{k}}/\|f+h\|)
\widehat{\sigma}_{t_{k}}(\chi_{\Omega})$ iff $b > \inf_{h \in {\mathfrak M}} \|f+h\| =
\|\hspace{.3em}[f]\hspace{.3em}\|$. But this is just $-b \widehat{\sigma}_{(f+
h)}(\chi_{\Omega}) \leq f+h \leq b \widehat{\sigma}_{(f+h)}(\chi_{\Omega})$.  Hence, $- b e
\leq [f] \leq b e$ for any $b \geq \|\hspace{.3em}[f]\hspace{.3em}\|$. Then for all $[f]
\in {\mathfrak W}^{\infty}, $ $-\|\hspace{.3em}[f]\hspace{.3em}\| e \leq [f] \leq
\|\hspace{.3em}[f]\hspace{.3em}\| e$.  It follows that the order interval $[- e, e]$ is
absorbing, and moreover, that $\|\hspace{.3em}[f]\hspace{.3em}\| \leq 1$ implies that $- e
\leq [f] \leq e$. The conclusion that $[-e,e]$ is an order ideal then follows immediately.
\qd

\vspace{3mm} {\bf 3. The states on $\winf$.}  Since the theory's Segal algebra is the
completion of  $\winf$, the states on $\winf$, denoted $\states$, will be identifiable
with the algebraic states. They may be characterized as follows. We give a second
characterization of them in terms of an order-unit norm below (Proposition III.14).

\begin{prop} Let $\phi \in {\mathcal K}(\winf)$.  Then $\| \phi \| = \phi(e) = 1$.
\end{prop}
\noindent \prf Note that for all $ [f] \in \winf$, if $\|[f]\| \leq 1$, then $-e \leq [f]
\leq e$.  Hence, if $\phi \geq 0$, then $\| \phi \| = \sup_{\|[f]\| \leq 1} |\phi([f])|
\leq \phi(e) = 1$.  But $\| \widehat{\sigma_t}(\chi_{\Omega}) \| = \| \chi_{\Omega} \|_t =
1.$  Then since the canonical surjection $\tau$ is a contraction, $\|
\tau(\widehat{\sigma_t}(\chi_{\Omega})) \| = \| e \| \leq 1.$  Hence, $\phi(e) \leq \| \phi
\|$, and therefore $\| \phi \| = 1$. \qd

\noindent The wk*-compactness of \states is similar to that shown in  Proposition II.1.

\vspace{3mm} {\bf 4. The Kadison representation of \winf .}   The functional
representation is directed by the requirements of Kadison's theorem \cite{kadi51}. We
must begin with the most basic properties of the order on $\winf$. Denote by $C$ the
positive cone of $\bigvee_1 \wat{t}$, {\em i.e.}, the set of all nonnegative elements,
and by $[C]$ the positive cone in $\winf$. The general properties of the order relation
$\leq$ in $\winf$ are given in Lemma III.3.  These do not assure that the quotient order
is antisymmetric \cite[2.3.4]{sema71}. Since antisymmetry is needed in the functional
representation of $\winf$, it must therefore be shown directly.  The proof will depend on
the following lemma.

\begin{lemma} The only element $h \in {\mathfrak M}$ comparable to 0 is $h$
= 0 itself, i.e., $C \cap {\mathfrak M} = \{0\}.$
\end{lemma}

\noindent  {\em Proof.} Fix any $h = \sum_{k=1}^p \widehat{\sigma}_{s_k}h^{s_k} \in
{\mathfrak M}$, $p$ finite.  By definition, $h$ is a linear combination of pairs of the
form $\widehat{\sigma}_{s_k}f^{s_k} -
\widehat{\sigma}_{s_k^{\prime}}\widehat{\eta}^{s_k^{\prime}}_{s_k}f^{s_k}$, so we write
$\sum_{k=1}^p \widehat{\sigma}_{s_k}h^{s_k} = \sum_{k=1}^p (\widehat{\sigma}_{s_k}f^{s_k} -
\widehat{\sigma}_{s_k^{\prime}}\widehat{\eta}^{s_k^{\prime}}_{s_k}f^{s_k})$. Suppose $h <
0$. Let $s \geq s_k$ for all $k$, and denote $\varepsilon_k = \sup_{x \in
\Omega}h^{s_k}(x)$ and $\widehat{\eta}^s_{s_k}(\varepsilon_k) = \sup_{x \in
\Omega}\widehat{\eta}^s_{s_k}h^{s_k}(x).$ For some $k$, say $k=1$, $h^{s_k} < 0.$  Then
there exists $x \in \Omega, \epsilon > 0$ such that $\widehat{\eta}^s_{s_1}h^{s_1}(x) < -
\epsilon.$  Of course, $\widehat{\eta}^s_{s_k}h^{s_k}(x) \leq
\widehat{\eta}^s_{s_k}(\varepsilon_k)$ for all other $k$.  Writing out each component and
summing over the $p$ inequalities yields

\begin{equation}
0 < -\epsilon + \sum_{k = 2}^p
\widehat{\eta}^s_{s_k}(\varepsilon_k)
\end{equation}

\noindent The 0 on the left comes from the fact that the contribution from each pair
$\widehat{\sigma}_kf^{s_k}(x) -
\widehat{\sigma}_{s_{k^{\prime}}}\widehat{\eta}^s_{s_k^{{\prime}}}f^{s_k}(x)$ is just
$\widehat{\eta}^s_{s_k} f^{s_k}(x) -
\widehat{\eta}_{s_k^{{\prime}}}^s(\widehat{\eta}^{s_k^{{\prime}}}_{s_k}f^{s_k})(x) =
\widehat{\eta}^s_{s_k}f^{s_k}(x) - \widehat{\eta}^s_{s_k}f^{s_k}(x) = 0.$  From this
equation, there exists at least one $k \neq 1$ such that
$\widehat{\eta}^s_{s_k}(\varepsilon_k) > 0.$  Then $h^{s_k} \vee 0 \geq
\widehat{\eta}^s_{s_k} h^{s_k} \vee 0 > 0.$  But this is impossible if $h < 0.$ The proof
for $h > 0$ is similar.

The countable case is simplified by the fact that the $l_1$-join $\bigvee_1 \wat{t}$ must
be a Banach space, and in particular, that the norm $\| h \| = \sum_{k=1}^{\infty} \| h^k
\|_k < \infty$. With sup norm, this means that for any choice of $ \epsilon$ in eq. (3.2),
the positive contribution must come from the first $p$ terms for $p$ sufficiently large.
The above proof therefore applies here as well. We need to show the existence of the
summations. For any function $h \in {\mathfrak M}$, $h =
\sum_{k=1}^{\infty}\widehat{\sigma}_{s_k}h^{s_k}$, $\sum_{k=1}^{\infty} |\varepsilon_k|
\leq \sum_{k=1}^{\infty} \| h^{s_k} \|_{s_k} = \| h \| < \infty$, and since
$|\widehat{\eta}^s_{s_k}(\varepsilon_k)| \leq |\varepsilon_k|$, $\sum_{k=1}^{\infty}
|\widehat{\eta}^s_{s_k}(\varepsilon_k)| \leq \infty$.  \qd

The antisymmetry of the quotient order then follows immediately. It is displayed
here together with two other important (and actually equivalent)
properties involving the order.

\begin{prop}
The following properties obtain:

{
\renewcommand{\labelenumi}{(\roman{enumi})}
\begin{enumerate}
\item The quotient set ${\mathfrak M}$ is an order ideal;
\item The positive cone $[C]$ is proper; and
\item The quotient partial order $\leq$ is antisymmetric.
\end{enumerate}
}

\end{prop}
\noindent {\em Proof.} For (i), ${\mathfrak M}$ is an {\bf order
ideal} iff for any pair $h_{1}, h_{2} \in {\mathfrak  M}, h_{1}
\leq g \leq h_{2}$ implies $g \in {\mathfrak M}$.  But $0 \leq g -
h_{1} \leq h_{2}- h_{1}$ implies by the lemma that $h_{2} - h_{1}
= 0$, or $g = h_{1} \in {\mathfrak M}$.

For (ii), suppose $[g] \in [C] \bigcap (-[C])$  The cone $[C]$ is
said to be {\bf proper} iff this implies that $[g] = 0$.  Suppose
$[g] \leq [0] \leq [g]$. The two inequalities require that there
exist $h_{1}, h_{2} \in {\mathfrak M}$ such that $g \leq h_{1}$
and $0 \leq g + h_{2}$.  Then $0 \leq g + h_{2} \leq h_{1} +
h_{2}$.  But from the lemma, $0 \leq h_{1}+h_{2}$ implies that
$h_{1}+h_{2} = 0$, or $0 \leq g + h_{2} \leq 0$.  Then by the
partial order on ${\bf \bigvee_{1}}{\mathfrak W}({\mathfrak
A}^{t}), g + h_{2} = 0$, or $g = -h_{2} \in {\mathfrak M}$.
Hence, $[g] = [0]$.

For (iii), if $[f] \leq [g] \leq [f]$, then $[0] \leq [g] - [f]
\leq [0]$. Hence, from (ii), $[g - f] = [0]$, or $[g] = [f]$.\qd

\noindent In the following, the term {\em order} will always imply
the antisymmetric property.

Although $\winf$  is by definition a Banach space with respect to its quotient norm, its
representation in $\C$ will be based instead on a norm which makes direct use of its order
unit $e$.  Denote by $E$ the order interval $[-e,e] = \{[f]: -e \leq [f] \leq e \}$.

\begin{prop}
The Minkowski functional $p_{E}([f]) = \inf\{b>0: -be \leq [f]
\leq b e$\} is a continuous seminorm on the Banach space
${\mathfrak W}^{\infty}$, and $p_{E} \leq \|\makebox[1em]{.}\|.$
\end{prop}

\noindent {\em Proof.} Using the fact that ${\mathfrak M}$ is a linear subspace, one
readily shows that the order interval $E$ is a convex, balanced, and absorbing set in
$\winf$.  But the Minkowski functional of any such set is a seminorm. Clearly $p_{E} \leq
\|\makebox[1em]{.}\|,$ and hence $p_{E}$ is $\|\makebox[1em]{.}\|$-continuous.\qd

The seminorm $p_E$ is a norm on $\winf$ if  $\winf$  is Archimedean \cite[II.1.2] {alfs71}.
This result is assured by the next proposition.

\begin{prop}
The Banach space ${\mathfrak W}^{\infty}$ with its quotient
partial order is Archimedean.  The positive cone $[C]$ is
$\|\makebox[1em]{.}\|$-closed, and ${\mathfrak W}^{\infty} = [C] -
[C],$ i.e., $[C]$ is generating.
\end{prop}
\noindent {\em Proof.} We show first that the order interval $[-e,
e]$ is $\|\makebox[1em]{.}\|$-closed.  Fix any Cauchy sequence
$([f_{n}])_{n \in {\mathbf N}} \in [-e, e].$  Since ${\mathfrak
W}^{\infty}$ is complete, the limit $[f]$ exists in ${\mathfrak
W}^{\infty}$.  We claim that $[f] \in [-e, e]$.  In fact, $0 \leq
|p_{E}([f_{n}]) - p_{E}( [f])| \leq p_{E}([f_{n}] - [f]) \leq
\|\hspace{.2em}[f_{n}] - [f]\hspace{.2em}\| \rightarrow 0$.  Then
$p_{E}([f]) = \lim p_{E}([f_{n}]) \leq 1$, because $[f_{n}] \in
[-e, e]$ implies that $p_{E}([f_{n}]) \leq 1\hspace{.3em}\forall n
\in$ {\bf {\sf N}}.  But $p_{E}([f]) \leq 1$ implies that $[f] \in
[-e, e]$.  It follows immediately that $[-a e, a e]$ is closed for
any $a > 0$.  Now let $([f_{n}])_{n \in {\mathbf N}} \in [C]$ be
any $\|\makebox[1em]{.}\|$-Cauchy sequence in the positive cone
$[C]$, and let $\lim[f_{n}] = [f]$. We show that $[f] \in [C]$.
Fix any $\varepsilon > 0.$  Then there exists $n_{\circ} \in$ {\bf
{\sf N}} such that for all $n \geq n_{\circ} ,
|\hspace{.2em}\|\hspace{.2em}[f_{n}]\hspace{.2em}\|
-\|\hspace{.2em}[f_{n_{\circ}}] \hspace {.2em}\|\hspace{.3em}|
\leq \varepsilon.$ Now fix any $a \geq
(1/2)(\|\hspace{.3em}[f_{n_{\circ}}]\hspace{.3em}\| +
\varepsilon)$, so that $[0] \leq [f_{n}] \leq
\|\hspace{.3em}[f_{n}]\hspace{.3em}\| e \leq 2 a e \hspace{.5em}
\forall n \geq n_{\circ}.$ The set $[\hspace{.3em}[0], 2 a
e\hspace{.3em}]  = a e + [-a e, a e]$ is a closed neighborhood of
$a e$, so that $[f] \in [\hspace{.3em}[0], 2a e\hspace{.3em}]$.
But $[\hspace{.3em}[0], 2ae\hspace{.3em}] \subset [C]$.

The proof of the Archimedean property follows directly.  Suppose
$n[f] \leq e\hspace{.5em}\forall n \in$ {\bf {\sf N}}.  We must
show that $[f] \leq [0].$  Since $(1/n)e \rightarrow [0],$ and
$(1/n)e - [f] \in [C]$ (closed), $ \lim((1/n)e -[f]) = -[f] \in
[C].$ To show that the cone $[C]$ is generating, note simply that
for every $[f] \in {\mathfrak W}^{\infty}, [0], [f] \leq
\|\hspace{.3em}[f]\hspace{.3em}\|e.$  Then $[f] =
\|\hspace{.3em}[f]\hspace{.3em}\| e -
(\|\hspace{.3em}[f]\hspace{.3em}\| e - [f]) \in [C] -[C]$.\qd

With the change in norm on $\winf$, it is useful to introduce a new norm on the dual
$(\winf)$* as well.     Define $\|\makebox[1em]{.}\|_{p}$ on $({\mathfrak W}^{\infty})$*
by $\|\phi\|_{p} = \sup_{p_{E}([f]) \leq 1} |\phi([f])|.$ We may characterize the set of
states $\states$ in terms of the new norm  as follows.

\begin{prop}
The linear functional $\phi \in (\winf)$* is  a state on  $\winf$  iff $\| \phi \|_{p} =
\phi(e) = 1.$
\end{prop}
\noindent
{\em Proof.}
Suppose $\phi \in \states$.  Since $p_{E}([f])
\leq 1$ implies that $-e \leq [f] \leq e, |\phi([f])| \leq
\phi(e) = 1$ from $\phi \geq 0,$ and hence $\| \phi \|_{p} =1.$
Conversely, suppose $\|\phi\|_{p} = \phi(e) = 1.$  We must show
that $\phi \geq 0.$  If $[0] \leq [f] \leq e,$ then $[0] \leq e
- [f] \leq e,$ and therefore $p_{E}( e - [f]) \leq 1.$  Then
$|\phi(e) - \phi([f]) | \leq |\phi(e - [f])| \leq \|\phi \|_{p}
= 1,$ and hence $|1 - \phi([f])| \leq 1.$  Then $\phi([f]) \geq 0.$
\qd

The representation of $\winf$ now follows immediately from Kadison's functional
representation of an order-unit space.

\begin{thm}
Let $\W_K = {\mathcal A}(K)$ be the Banach space of continuous affine functions on the \cc
set $K$ of states on  $\winf$. The linear space $({\mathfrak W}^{\infty}, e)$ with order
unit $e$ and norm $p_E$ has a functional representation $\Delta_K:\winf \rightarrow
{\mathfrak W}_K$ defined by $\Delta_K ([f])(\phi) = \phi([f]).$  The function $\Delta_K$ is
a $p_{E}$-isometry, order-preserving in both directions. The image $\Delta_K(\winf)$ is a
separating uniformly dense subset of the Banach space ${\mathfrak W}_K$ and $\Delta_K(e)$
is the constant function $1_K$ on $K$.
\end{thm}

\noindent {\em Proof.}  \cite{kadi51}, \cite[Theorem
II.2.9]{alfs71}\hspace{5mm}\qd

\noindent The space $\W_K$ is the uniform closure of the subspace $\Delta_K(\winf)
\subset {\mathcal C}(K)$.  It is therefore a completion of $\winf$ with respect to the
order-unit norm $p_E$. We refer to it throughout as {\em the} completion of $\winf$.
Denote its elements by $\widehat{f} \in \W_K$.

\begin{center} {\bf  D. The $MI$-spaces of observables.}
\end{center}

Much of the theory's quasilocal structure depends on the fact that the space $\W_K$ is an
$MI$-space.  We now prove this fact. In particular, this will provide a representation of
$\W_K$ as $\C$.

\begin{thm} The space ${\mathfrak W}_K$ is an $MI$-space.
\end{thm}

\noindent {\em Proof.} We show first that the union $\bigcup_{t \in {\bf J}} \wat{t}$ maps
to a uniformly dense subspace of $\winf$ under the injections $(\sigma_t)$.  Fix any $[f]
\in \winf$, and any $f \in \tau^{\leftarrow}([f]) \in \bigvee_1 \wat{t}$. For some
countable set of indices $(t_k) \in {\bf J}$, $f = \sum_{k=1}^{\infty}
\widehat{\sigma}_{s_k}f^{s_k}$.  For any given $\varepsilon > 0$, there exists $t_o \in
{\bf N}$ such that for all $t > t_o$, $\|f - \sum_{k = 1}^t \sigma_{s_k}f^{s_k} \| <
\varepsilon.$ From the composition rule, $\sigma_s = \sigma_t\ehat$, so that $\sum_{k=1}^t
\sigma_{s_k}f^{s_k} = \sum_{k=1}^t \sigma_t\widehat{\eta}_{s_k}^tf^{s_k} =
\sigma_t\sum_{k=1}^t \widehat{\eta}_{s_k}^tf^{s_k}.$  Write $g^t = \sum_{k=1}^t
\widehat{\eta}_{s_k}^tf^{s_k}.$ By the definition of the quotient norm, $\|[f] - [g^t]\| <
\varepsilon$.

We can readily show that the image $\tau  (\bigcup_{t \in {\bf J}}
\widehat{\sigma}_t\wat{t})$ is a vector lattice.  For given any pair $s,t \in {\bf J}$,
fix $f^s \in \wat{s}$, $g^t \in \wat{t}$. Since ${\bf J}$ is upward directed, there
exists a $u \in {\bf J}$ such that $s,t \leq u$.  From the composition rule, $\sigma_sf^s
= \sigma_u\widehat{\eta}_s^uf^s$ and $\sigma_tg^t = \sigma_u\widehat{\eta}_t^ug^t$.  Then
$\widehat{\eta}_s^uf^s, \widehat{\eta}_t^ug^t \in \wat{u}$ (a Banach lattice), so that
$\widehat{\sigma}_u(\widehat{\eta}_s^uf^s \vee \widehat{\eta}_t^ug^t) =
\widehat{\sigma}_u\widehat{\eta}_s^uf^s \vee \widehat{\sigma}_u\widehat{\eta}_t^ug^t$,
because $\widehat{\sigma}_u$ is the natural injection.  Since the surjection $\tau:
\bigvee_1 \wat{t} \rightarrow \winf$ is order-preserving (Lemma III.3),
$\sigma_u(\widehat{\eta}_s^uf^s \vee \widehat{\eta}_t^ug^t) =
\sigma_u\widehat{\eta}_s^uf^s \vee  \sigma_u\widehat{\eta}_t^ug^t = \sigma_sf^s \vee
\sigma_tg^t$. That is, $[f^s] \vee [g^t] = [\widehat{\eta}_s^uf^s \vee
\widehat{\eta}_t^ug^t] \in \winf$. Furthermore, the subspace $\Delta_K \tau  (\bigcup_{t
\in {\bf J}} \widehat{\sigma}_t\wat{t})$ is uniformly dense in ${\mathfrak W}_K$ because
$\Delta_K$(\winf) is dense.  Note especially that for any $f \in \join$, $p_E  \leq
\|[f]\| \leq \|f\|$ from Proposition III.9 and the properties of the quotient norm on
$\winf$.

We show that the mappings $\Delta_K \circ \sigma_t: \wat{t} \rightarrow {\mathfrak W}_K$
are 1:1. Recall first that for any $x \in \Omega$, the Dirac point functional $\delta(x)$
defined by $\delta(x)(f^t) \equiv f^t(x)$ is a state on \wat{t}, i.e., $\delta(x) \in E_t$,
for all $t \in$ {\bf J}. Furthermore, $\Delta_K \circ \sigma_t(f^t)(\phi_{\mu}) =
\phi_{\mu}(\sigma_tf^t) = \mu_t(f^t)$. Now suppose $f^t \neq g^t$. For some $x \in \Omega$,
$\delta(x)(f^t) \neq \delta(x)(g^t).$ Then $\Delta_K \circ \sigma_t(f^t)(\delta(x)) \neq
\Delta_K \circ \sigma_t(g^t)(\delta(x))$. Thus, $\Delta_K \circ \sigma_t(f^t) \neq \Delta_K
\circ \sigma_t(g^t).$

We show that for any $t \in {\mathbf J}$, $\Delta_K\circ \sigma_t$ is order-preserving in
both directions.  Fix $f^t \geq g^t$.  Then $\forall \mu \in E_{\infty}$, $\mu_t(f^t - g^t)
\geq 0$, and therefore $\Delta_K\circ\sigma_t(f^t - g^t)(\phi_{\mu}) \geq 0$. Hence,
$\Delta_K\circ\sigma_t(f^t) \geq \Delta_K\circ\sigma_t(g^t).$  Conversely, suppose
$\Delta_K\circ \sigma_t(f^t) \geq \Delta_K\circ\sigma_t(g^t)$.  Then for all $\mu_t \in
E_t$, $\mu_t(f^t) \geq \mu_t(g^t)$, and in particular $\delta(x)(f^t) \geq \delta(x)(g^t)
\hspace{2mm}\forall x \in \Omega$.  Hence, $f^t(x) \geq g^t(x)\hspace{2mm}\forall x \in
\Omega$, and therefore $f^t \geq g^t$.  It follows that $\Delta_K\circ\sigma_t$ is a
lattice homomorphism.

We have thus shown that the image  $\tau  (\bigcup_{t \in {\bf J}}
\widehat{\sigma}_t\wat{t})$ is a uniformly dense linear subspace of ${\mathfrak W}_K$   and
a normed vector lattice.  Then its closure  ${\mathfrak W}_K$ is a Banach lattice. The
constant function $1_K \equiv 1 \in {\mathfrak W}_K$ is an order unit in $\W_K$, {\em
i.e.,} $\|\widehat{f}\| \leq 1$ iff $|\widehat{f}| \leq 1_K$. Then ${\mathfrak W}_K$ is an
$MI$-space (\cite{sema71}, Proposition 13.2.4).\qd

The space ${\mathfrak W}_K = {\mathcal A}(K)$, with $K = \states$, is the (essentially
unique) order-unit completion of $\winf$. We take it as the theory's space of quasilocal
observables as required by Axiom 4. We now generalize to allow a choice of $K$.

\begin{cor}
Let $K \subseteq {\mathcal K}(\winf)$ be any nonempty compact convex set of states.  Then
the space ${\mathfrak W}_K = {\mathcal A}(K)$ of continuous affine functions on $K$ is an
$MI$-space. The Kadison function $\Delta_K: \winf \rightarrow {\mathfrak W}_K$ is an
order-preserving mapping onto a dense subset of $\W_K$, and the order unit $1_K \in
{\mathfrak W}_K$.
\end{cor}

\noindent {\em Proof.} Let ${\mathcal A}(K; {\mathcal K}(\winf))$ be the set of functions
${\mathcal A}(\states)$ restricted to $K$. ${\mathcal A}({K; \states})$ is a uniformly
dense subset of ${\mathcal A}(K)$ (\cite[Corollary I.1.5]{alfs71}, \cite[23.3.6]{sema71}).
For the restrictions, note that for all $\widehat{f}, \widehat{g} \in {\mathcal
A}(\states)$, $\widehat{f} \leq \widehat{g} \Rightarrow \widehat[f]|_K \leq
\widehat[g]|_K.$ In particular, $|\widehat{f}| \leq [e] \Rightarrow |\widehat{f}|_K \leq
[e]|_K.$ By the theorem, ${\mathcal A}({\mathcal K}(\winf))$is a Banach lattice, so that
${\mathcal A}(K; {\mathcal K}(\winf))$ is a normed vector lattice. Then ${\mathfrak W}_K$
is the completion of a normed vector lattice, and therefore, a Banach lattice
\cite[Proposition 3.9.5]{sema71}. Since there can be no confusion in context, we shall also
write $\widehat{f} \in \W_K$ to denote its elements.\qd

\spg Henceforth,  $K \subseteq \states$ will always denote an arbitrary \cc set of states.

\vspace{3mm}
\begin{center} {\bf E. Representation in ${\mathcal C}(X_K)$.}   \end{center}

Since $(\W_K, {\bf 1}_K)$ is a partially ordered Banach space with unit, we may
characterize its states as follows.

\begin{prop} The  states on $\W_K$, denoted ${\mathcal K}(\W_K)$, are a compact set consisting of the
positive linear functionals on $\W_K$ for which $\|\phi\| = \phi(1_K) = 1.$
\end{prop}

\noindent \emph{Proof.} The proof is similar to that in Proposition II.1. As the
intersection of a wk*-closed hyperplane and the compact unit ball $\bigcirc$*$\W_K$ in the
(topological) dual $\W_K$* of $\W_K$, ${\mathcal K}(\W_K)$ is compact.   \vspace{5mm} \qd

The states on $\W_K$  are related to those on $\winf$ by the following.

\begin{prop} Let $K \subseteq \states$ be any \cc set, and
define $\alpha_K: K \rightarrow \mathcal{K}(\W_K)$ by $\alpha_K \phi_{\mu}(\widehat{f}) =
\widehat{f}(\phi_{\mu}) $. Then $\alpha_K$ is an affine homeomorphism giving a
parameterization or {\em indexing}   of $\mathcal{K}(\W_K)$ by $x_{\mu} =
\alpha_K(\phi_{\mu}).$
\end{prop}

\noindent {\em Proof.}  \cite[Theorem 23.2.3]{sema71}.  For the affine property, one has
that for all $\mu, \nu \in K$ and $a \in (0,1)$, $\alpha_K(a \mu + (1-a)
\nu)(\widehat{f}) = \widehat{f}(a \mu + (1- a) \nu) = a \widehat{f}(\mu) +
(1-a)\widehat{f}(\nu) = (a \alpha_K \mu + (1-a)\alpha_K\nu)(\widehat{f})$.
\hspace{1cm}\qd

\noindent Note the dependence on choice of $K \subset {\mathcal K}(\winf)$.

\begin{cor} The set of extremal states $\partial_e{\mathcal K}(\W_K)$ is closed and
therefore compact. \end{cor}

\noindent {\em Proof.} Clearly, $\alpha_K(\partial_eK) =
\partial_e\mathcal{K}(\W_K).$  But ${\mathfrak W_K} = {\mathcal
A}(K)$ is a vector lattice.  Hence, the set of states $K$ is a {\em regular (or Bauer)
simplex}, {\em i.e.,} a simplex for which the set of extremal points $\partial_eK$ is
wk*-closed (\cite[23.7.1]{sema71}, \cite{baue61} ). \qd

We are now able to define the algebra $\mathcal{C}(X_K)$.

\begin{thm}
Let $X_K = \partial_e{\mathcal K}(\W_K)$. The mapping $\psi_K: {\mathfrak W_K} \rightarrow
{\mathcal C}(X_K)$ defined by $\psi_K(\widehat{f})(x_{\mu}) = x_{\mu}(\widehat{f})$ is an
isometric vector-lattice isomorphism onto ${\mathcal C}(X_K)$ with $\psi_K(1_K) = 1_{X_K}.$
\end{thm}

\noindent {\em Proof.} Apply the structure theorem for $MI$-spaces
to the pair $(\W_K, 1_K)$ \cite[Theorems 13.2.3, 13.2.4]{sema71}.
\qd

The $MI$ spaces satisfy all the linear postulates of Segal algebra, but they do not have a
vector multiplication needed to define powers.  The isomorphism with $\C$ permits us to
assign the operation as follows.

\begin{prop} Define vector multiplication on $\W_K$ by $\widehat{f}\cdot \widehat{g} =
\psi_K^{-1}(\psi_K(\widehat{f}) \cdot \psi_K(\widehat{g})$ for all $\widehat{f},
\widehat{g} \in \W_K$. Then $\W_K$ is a Segal algebra.  \qd   \end{prop}

\spg This completes the requirement of Axiom 4.  In most of what follows, however, the
representation of ${\mathfrak W}_K$ as $\C$ will be found to play the major role.

\vspace{3mm} \begin{center} {\bf E. Choquet decompositions.}  \end{center}

Compact convex sets of states play an important role in the modern theory of  statistical
mechanics.  The theory of this class of states depends crucially on the unique
decomposition of states into extremal (or {\em pure}) states.  We now show that this
result is assured by the fact  that the set of extremal states $\partial_e{\mathcal
K}(\wo)$ is closed (Corollary III.16).

Let $\partial_eK$ be the set of extremal points of $K \subseteq \states$, and ${\mathcal
S}(\partial_eK)$ the set of Radon probability measures on $\partial_eK$ with the topology
induced on it by the wk*-topology on $K$ under the Riesz representation theorem
${\mathcal S}(\partial_eK) = {\mathcal K}{\mathcal C}(\partial_eK).$

\begin{thm} Let $K \subseteq \states$ be any \cc set of states.  Then its set of extremal
states $\partial_eK$ is closed. Hence, for each state $\phi_{\mu} \in K$, there exists a
unique Radon probability measure $\sigma_{\mu}^{\prime}$ on $K$ with
$\sigma_{\mu}^{\prime}(\partial_eK) = 1$ such that

\begin{equation}
\widehat{f}(\phi_{\mu}) = \int_{\partial_eK} \widehat{f}(\phi)
d\sigma_{\mu}^{\prime}(\phi)\hspace{1cm}\forall \widehat{f} \in {\mathfrak W_K}
\end{equation}

\noindent Let ${\mathfrak r}: K \rightarrow {\mathcal S}(\partial_eK)$ map states to the
corresponding probability measures, i.e., ${\mathfrak r}(\phi_{\mu}) =
\sigma_{\mu}^{\prime}$. Then ${\mathfrak r}$ is an affine homeomorphism onto ${\mathcal
S}(\partial_eK)$.
\end{thm}

\noindent {\em Proof.} (\cite[Theorem II.4.1]{alfs71}).  The fact that $\partial_eK$ is
closed (by Corollary III.16) assures the existence of the measure $\sigma_{\mu}^{\prime}$
\cite[23.4.8]{sema71}. The fact that $K$ is a simplex (by the proof of the same
corollary) assures the uniqueness \cite[23.6.5]{sema71}.\hspace{1cm}\qd

\spg  Integrals of this form are called the {\em Choquet decomposition} (\cite{choq66},
\cite{phel66})  of the  given state $\phi_{\mu} \in K \subseteq {\mathcal K}({\winf})$
into the set of pure states $\partial_eK$. A state $\phi_{\mu}$ satisfying this equation
is called the {\em centroid} (or {\em resultant}) of the probability measure
$\sigma^{\prime}_{\mu}.$

 \vspace{5mm} \noindent{\bf {\large IV \hspace{4mm} Indexing of states }}
\setcounter{section}{4} \setcounter{equation}{0} \setcounter{thm}{0}

\vspace{3mm} Up to this point, we have seen two kinds of states.  The one arises through
the definition of the texture itself (Axiom 1), the mapping $\Lambda_t \mapsto \wat{t}$.
We immediately defined the corresponding state space $E_t = {\mathcal K}\wat{t}$
(Proposition II.1) of local states $\mu_t \in E_t$ such that $\mu_t(f^t)$ is the
expectation value of the observable $f^t \in \wat{t}$ if the lattice is in state $\mu_t$.
The other kind of state is the global state $\phi_{\mu} \in {\mathcal K}(\wo)$ on the
Segal algebra $\wo$. For every global observable $\widehat{f} \in \wo$, the expectation
value is $\phi_{\mu}(\widehat{f})$. We know how to map any local observable $f^t$ to its
global representation $\widehat{f}$.  In this section, we learn that every local state
$\mu_t$ has a unique global representation and how to identify it.

We shall prove by traditional methods the following property of the category-theoretical
limits of the states of the theory.  It can be shown that the functor ${\mathcal K}$,
which maps the $MI$-spaces $\wat{t}$ to their sets of states ${\mathcal K}(\wat{t})$ and
the morphisms $\ehat:\wat{s} \rightarrow \wat{t}$ to the ${\mathcal K}(\ehat) =\enohat:
E_t \rightarrow E_s$,  is a {\em directly continuous functor}, {\em i.e.,} it maps
inductive limits to projective limits according to the rule

\begin{equation} {\mathcal K}(\mbox{$\lim^{\rightarrow}$}(\dirsys)) = \mbox{$\lim^{\leftarrow}$}(\{{\mathcal
K}(\wat{t}, \enohat, {\mathbf J}\})
\end{equation}

\spg  The left-hand side gives the theory's algebraic states. The right-hand side depends
on the system $\{{\mathcal K}(\wat{t}, \enohat, {\mathbf J}\} \equiv \{E_t, \enohat,
{\mathbf J}\}$, which  has already been introduced following Proposition III.1 ({\em cf.}
\cite[11.8.6 and 23.3.2]{sema71}).

\vspace{3mm}\begin{center}  {\bf  A. The limit $\{E_{\infty}, \rho_t,$ $ {\mathbf J}\}$}.
\end{center}

We begin by constructing the projective limit.

\begin{prop}
The system $\{E_t, \enohat, {\mathbf J}\}$ has a unique {\bf C}{\em ompconv}-projective
limit $\{E_{\infty}, \rho_t,$ $ {\mathbf J}\} = {\mbox {{\em lim}}}^{\leftarrow}\{E_t,
\enohat, {\mathbf J}\}$, with nonempty compact object set $E_{\infty}$ and continuous
affine mappings $\rho_{t}: E_{\infty} \rightarrow E_{t}$ obeying the composition rule
$\rho_{s} = \eta^{t}_{s} \rho_{t}$ for all $t \geq s, s \in {\mathbf J}$.
\end{prop}

\noindent \prf The transformations $(\eta^{t}_{s})_{s \leq t}$ obey the set of composition
rules required to make the set $\{ E_{t}, \eta^{t}_{s}, \mathbf{ J} \}$ a projective system
of compact convex spaces. The typical commuting diagram is as follows, $\forall r \leq s
\leq t$:

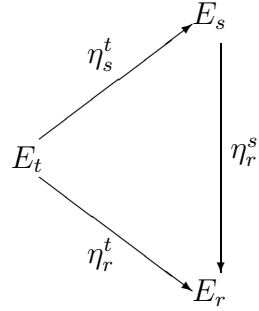
\setlength{\unitlength}{.01in}
\begin{picture}(250,200)(-140,0)
\put(45,90){$E_{t}$} \put(140,20){$E_{r}$} \put(140,165){$E_{s}$}
\put(85,40){$\eta^{t}_{r}$} \put(160,95){$\eta^{s}_{r}$} \put(85,145){$\eta^{t}_{s}$}
\put(60,85){\vector(4,-3){80}} \put(60,105){\vector(4,3){80}}
\put(155,155){\vector(0,-1){120}}
\end{picture}

\noindent A proof from traditional topology then applies, as given, {\em e.g.}, in
\cite[Theorem 3.2.10]{enge68}. The construction in a category-theoretical setting is
provided by \cite[Proposition 11.8.2]{sema71}. The limit object  is the compact subspace of
the Cartesian product of the $(E_{t})_{s \leq t}$ defined by

\[
\mbox{lim}^{\leftarrow} \{E_t, \enohat, {\mathbf J} \} = \{(\mu_{t})_{t \in \mathbf{J}} \in
{\mbox{{\bf{\sf P}}}}{}_{t \in {\mathbf J}} E_{t} : \mu_{s}= \enohat
\mu_t\hspace{1em}\forall s \leq t,\hspace{.5em}s,t \in \mathbf {J}\}
\]

\noindent Denote the limit object by $E_{\infty}$. The limit morphisms $\rho_t: E_{\infty}
\rightarrow E_t$ are defined by $\rho_t((\mu_s)_{s \in {\mathbf J}}) = \mu_t$.\qd

\spg The elements of $E_{\infty}$ are commonly called {\bf threads}.

\begin{prop}
The mappings $\rho_{t}: E_{\infty} \rightarrow E_{t}$ are injective, i.e., for any state
$\mu_{t} \in \rho_{t}(E_{\infty})$ there exists exactly one thread $(\mu_{t})_{t \in
{\mathbf J}} \in E_{\infty}$  such that $\rho_{t}(\mu) = \mu_{t}$.  Moreover, the
transformations $\eta^{t}_{s} : E_{t} \rightarrow E_{s}$ map the set $\rho_{t}(E_{\infty})
\subseteq E_{t}$ 1:1 onto the set $\rho_{s}(E_{\infty}) \subseteq E_{s}$.
\end{prop}
\noindent \prf Take the second part of the proposition first.  To show that $\eta^{t}_{s}$
is onto $\rho_{s}(E_{\infty})$, fix any $\mu_{s} = \rho_{s} \mu$.  Then $\rho_t \mu \in
E_t$, and $\eta^t_s \rho_t \mu = \mu_s.$ For $\eta^t_s$  1:1, fix any $s \leq t$, and
suppose that for some pair $\mu_t, \nu_t \in \rho_t(E_{\infty})$ , $\eta^t_s \mu_t =
\eta^t_s \nu_t$.  Then for all $f^s \in \wat{s}$, $\mu_t(\ehat f^s) =\nu_t(\ehat f^s).$  In
particular, for all $f^t \in \wat{t} \subseteq \wat{s}$, $\mu_t(\ehat f^t) = \mu_t(f^t)
=\nu_t(f^t),$ and therefore, $\mu_t = \nu_t$.  It follows that if $\mu_t \neq \nu_t,$ then
$\eta^t_s \mu_t \neq \eta^t_s \nu_t.$ To show that $\rho_{s}: E_{\infty} \rightarrow E_{s}$
is 1:1, fix any $\mu, \nu \in E_{\infty}, \mu \neq \nu.$  Then there exists $t \in {\mathbf
J}$ such that $\mu_t \neq \nu_t$.  Since the index set {\bf J} is upward directed, there
exists $u \geq s,t$.  Then $\eta^u_t \rho_u \mu =\rho_t \mu \neq \rho_t nu = \eta^u_t
\rho_u \nu$. Hence, $\rho_u\mu \neq \rho_u\nu.$  But $\eta^u_s$ is 1:1, so that $\mu_s =
\eta^u_s \mu_u \neq \eta^u_s\nu_u = \nu_s.$\qd

Let $r^{s}_{t}: E_{s} \rightarrow E_{t}$ be the restriction operator such that $r^{s}_{t}
\mu_{s} = \mu_{s}|_{\watt{t}}$. One then has from the Proposition that $r^{s}_{t}
\eta^{t}_{s}$ is the identity mapping $\iota_{E_{t}}$ on $E_t$.  This may be incorporated
into the following symmetrical form.

\begin{cor}
The operators $r^{s}_{t}$ and $\eta^{t}_{s}$ are mutually inverse on the subspaces
$\rho_{t}(E_{\infty})$ and $\rho_{s}(E_{\infty})$, i.e., for any comparable pair $s\leq t$,
\begin{displaymath}
     r^{s}_{t} \eta^{t}_{s} = \iota_{\rho_{t}(E_{\infty})}\hspace{5em} {and}\hspace{5em}\eta
^{t}_{s}  r^{s}_{t} = \iota_{\rho_{s}(E_{\infty})}
\end{displaymath}
\end{cor}
\noindent \prf We already have the one.  For every $\mu_{s} \in \rho_{s}(E_{\infty})$,
there exists a unique thread $\mu \in E_{\infty}$ such that $\mu_{s} = \rho_{s} \mu$.  Then
from the composition rule, $\eta^{t}_{s} r^{s}_{t} \mu_{s} = \eta^{t}_{s} r^{s}_{t}
(\eta^{t}_{s} \rho_{t} \mu) = \eta^{t}_{s}(r^{s}_{t} \eta^{t}_{s})\rho_{t} \mu =
\eta^{t}_{s} \rho_{t} \mu  = \mu_{s}$. \qd

Thus, higher components of a given thread are the restrictions of lower components, and the
lower components extensions of the higher. Both the $r^{s}_{t}$ and the $\eta^{t}_{s}$ map
every thread onto itself, {\em i.e.}, there is no mixing of threads under these
transformations. It follows immediately that    if a state $\mu_{t}$ for a particular
system $\Lambda_t$ belongs to a thread $\mu$ satisfying the condition $\mu_t = \rho_t \mu$,
then it determines that thread uniquely, and hence the state on every other system in the
space. Naturally, this mirrors thermodynamic equilibrium.

Now consider the defining condition for $\threads$  (Proposition IV.1):

\begin{equation}
\mu_s(f^s) = \mu_t(\ehat f^s) \hspace{7mm}\forall f^s \in \wat{s}
\end{equation}

\spg Since $\mu_s = \rho_s(\mu)$ and $\mu_t = \rho_t(\mu)$ derive from the same quasilocal
state $\mu \in \threads$, they represent the same state on their respective systems
$\Lambda_s \subseteq \Lambda_t$. In introduction of the morphisms $(\ehat)_{t \in {\bf
J}}$, it was assumed that they map local observables $f^s \in \wat{s}$ on the smaller space
$\Lambda^s$ to measurements $\ehat f^s \in \wat{t}$ of the same physical quantity on the
larger system $\Lambda_t \supseteq \Lambda_s$.   Eq. (4.2) is simply the requirement that
equivalent measurements on nested systems have the same expectation value.

\vspace{3mm} \begin{center} {\bf B. The homeomorphism \states = \threads}.  \end{center}

We now show $\states$ and $\threads$ are homeomorphic spaces, eq. (4.1), by conventional
means. This will provide the identification of the TL states to states on the Segal
algebras $\mathfrak{W}_K$. The proof depends on construction of a new projective limit of
compact spaces which is related to $E_{\infty}$. For all $t \in {\mathbf J}$, define the
mappings $\bigcirc$*$(\sigma_{t}): \bigcirc$*${\mathfrak W}^{\infty} \rightarrow
\bigcirc$*$\wat{t}$ by $\bigcirc$*$(\sigma_t) \phi(f^t) = \phi(\sigma_tf^t)$. (The
notation is again from category theory.) Since the injection $\sigma_t: \wat{t}
\rightarrow \winf$ is a linear contraction, the induced mapping $\bigcirc$*$(\sigma_t)$
is continuous. The mappings are in general into not onto. Fix any $t \in${\bf J}, and
since $E_t \subseteq \bigcirc$*$\wat{t}$, define $F_{\infty} =
\bigcirc$*$(\sigma_{t})^{\leftarrow}(E_{t})$. The set $F_{\infty}$ has the following
properties.

\begin{prop}
Let $F_{\infty} = \bigcirc$*$(\sigma_{t})^{\leftarrow}(E_{t})$ be the compact preimage in
$\bigcirc$*$ {\mathfrak W}^{\infty}$ of the state space $E_{t}$ for any $t \in \mathbf{J}$.
Then for any other $s \in \mathbf {J}, \bigcirc$*$(\sigma_{s})^{\leftarrow}(E_{s}) =
F_{\infty}$, i.e., $F_{\infty}$ is independent of choice of index. The  morphisms
$\bigcirc$*$(\sigma_{t})$ satisfy the composition rule $\bigcirc$*$(\sigma_{s}) =
\eta^{t}_{s} \bigcirc$*$(\sigma_{t})$ on $F_{\infty}$.
\end{prop}
\noindent \prf Since $E_t$ is compact, $F_{\infty}$ is a closed subset of the compact set
$\bigcirc$*${\mathfrak W}^{\infty}$, and therefore compact.  Note that on any subspace
$\sigma_s(\wat{s})$, the quotient surjection $\tau: \bigvee_{\mathbf 1} \wat{t} \rightarrow
\winf$ is contracting, since $\|\tau(\widehat{\sigma}_s f^s) \| = \| \sigma_s f^s \| =
\inf_{h \in {\mathfrak M}} \|f^s + h \| \leq \| f^s \| = \| \widehat{\sigma}_s f^s \|_s.$
Hence, $\|f^s \| \leq 1$ implies that $\| \sigma_s f^s \| \leq 1.$  In fact, $\|f^s \| \leq
1$ iff $\|\sigma_s f^s\|_s \equiv \| f^s \| \leq 1$, so that $\| \sigma_s f^s \| = \|
\tau(\widehat{\sigma}_sf^s \| \leq \| \widehat{\sigma}_s f^s \| \leq 1.$  To show that
$\bigcirc$*$(\sigma_s)^{\leftarrow} (E_s) \supseteq
\bigcirc$*$(\sigma_t)^{\leftarrow}(E_t),$ fix any $\phi \in
\bigcirc$*$(\sigma_t)^{\leftarrow}(E_t)$.  Then one has $\| \bigcirc$*$(\sigma_s) \phi \| =
\sup_{\| f^s \| \leq 1} | \bigcirc$*$(\sigma_s) \phi (f^s) | \leq \sup_{\| \sigma_s f^s \|
\leq 1} | \phi(\sigma_s f^s) |  \leq  \sup_{\| [f] \| \leq 1}  |\phi([f])| =  \| \phi \|
\leq 1$, because $\phi \in \bigcirc$*$(\winf)$.  But $\bigcirc$*$(\sigma_t)\phi \in E_t$
implies that $\bigcirc$*$(\sigma_t)\phi(\chi_{\Omega}) = \phi(\sigma_t \chi_{\Omega}) =
\phi(e) = 1.$ Then $\| \bigcirc$*$(\sigma_s)\phi \| = \bigcirc$* $(\sigma_s) \phi (1_{S_s})
= 1,$ and therefore, $\bigcirc$*$(\sigma_s) \phi$ is a state on \wat{s}, by Proposition
II.1. Show similarly that $\bigcirc$*$(\sigma_t)^{\leftarrow}(E_t) \supseteq
\bigcirc$*$(\sigma_s)^{\leftarrow}(E_s)$. Now for any comparable pair $s \leq t$, one has
that $\eta^{t}_{s}\bigcirc$*$(\sigma_{t})|_{F_{\infty}} =
\bigcirc$*$(\sigma_{s})|_{F_{\infty}}$, using the composition rule $ \enohat
\bigcirc$*$(\sigma_{t}) = \bigcirc$*$(\sigma_{t} \widehat{\eta}^{t}_{s})$, and then
$\bigcirc$*$(\sigma_{t} \widehat{\eta}^{t}_{s}) = \bigcirc$*$(\sigma_{s})$.\qd

\vspace{3mm} We show the equivalence of $F_{\infty}$ and $E_{\infty}$.

\begin{prop}

There exists a unique homeomorphic bijection $\beta: F_{\infty} \rightarrow E_{\infty}$
such that for any $t \in \mathbf{J}$, $\bigcirc$*$(\sigma_{t}) =\rho_{t} \beta. $ Then
$\{F_{\infty}, \bigcirc$*$(\sigma_{t}), {\mathbf J}\} = \lim^{\leftarrow}\{E_{t},$ $
\eta^{t}_{s}, {\mathbf J} \}$.

\end{prop}
\noindent \prf The composition rule $\bigcirc$*$(\sigma_{s}) = \eta^{t}_{s}
\bigcirc$*$(\sigma_{t})$ makes the following diagram commuting:

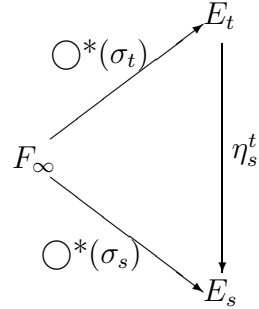
\setlength{\unitlength}{.01in}
\begin{picture}(250,200)(-140,0)
\put(40,90){$F_{\infty}$} \put(140,20){$E_{s}$} \put(140,165){$E_{t}$}
\put(55,40){$\bigcirc$*$(\sigma_{s})$} \put(155,95){$\eta^{t}_{s}$}
\put(60,145){$\bigcirc$*$(\sigma_{t})$} \put(60,85){\vector(4,-3){80}}
\put(60,105){\vector(4,3){80}} \put(150,155){\vector(0,-1){120}}
\end{picture}

\noindent From the uniqueness properties of the {\bf C}ompconv-projective limit $\{
E_{\infty}, \rho_t. \mathbf{J} \}$ \cite[11.8.1]{sema71}, one therefore has that there
exists a unique commuting homeomorphism $\beta:F_{\infty} \rightarrow E_{\infty}$
satisfying the composition rule $\bigcirc$*$(\sigma_{t}) = \rho_{t}
\beta\hspace{.5em}\forall t \in \mathbf{J}$:

\vspace{6mm} \setlength{\unitlength}{.01in}
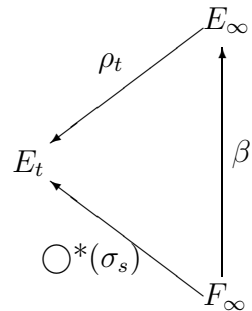
\begin{picture}(250,200)(-140,0)
\put(40,90){$E_{t}$} \put(140,20){$F_{\infty}$} \put(140,165){$E_{\infty}$}
\put(55,40){$\bigcirc$*$(\sigma_{s})$} \put(155,95){$\beta$} \put(85,145){$\rho_{t}$}
\put(140,25){\vector(-4,3){80}} \put(140,165){\vector(-4,-3){80}}
\put(150,35){\vector(0,1){120}}
\end{picture}

\noindent Moreover, if $\beta$ is onto, then $F_{\infty}$ is a new (but equivalent) {\bf
C}ompconv-projective limit of the system $\{E_t, \enohat, {\mathbf J}\}$.

Fix any $\mu = (\mu_t)_{t \in {\mathbf J}} \in E_{\infty}$, and define $\phi_{\mu}:
{\mathfrak W}^{\infty}\rightarrow$ \mbox{\bf \sf{R}} as follows.  For all $n \in$ \mbox{\bf
\sf{N}}, let $\phi_{\mu}(\hspace{.2em}\sum_{k = 1}^{n} \sigma_{t_k}(f^{t_k})\hspace{.2em})
= \sum^{n} \mu_{t_k}(f^{t_k})$ for any finite set of indices $(t_k) \in {\mathbf J}$. If
$[f] = \lim_{n \rightarrow \infty} \sum^{n} \sigma_{t_k}(f^{t_k}), $ let $\phi_{\mu}([f]) =
\lim_{n \rightarrow \infty} \sum^{n} \mu_{t_k}(f^{t_k})$. The limit exits for all $[f] \in
\winf$ because $|\sum \mu_t(f^t)| \leq \sum |\mu_t(f^t)| \leq \sum \|f^t \|_t  < \infty.$
({\em cf.} Proof, Theorem III.3.)  We seek to define $\alpha: E_{\infty} \rightarrow
F_{\infty}$ by $\alpha{\mu} = \phi_{\mu}$.

Clearly, the definition of $\phi_{\mu}$ assures that it is linear. In order for
$\phi_{\mu}$ to be well-defined on ${\mathfrak W}^{\infty}$, it must be independent of the
choice of representation of $[f]$. Since $[f] = [g]$ implies that $[f] - [g] = [f - g] =
[0],$ it suffices to show that $\phi_{\mu}([0]) = 0.$ The preimage of $[0]$ under the
quotient surjection is the set ${\mathfrak M}$, the closed linear span of elements of ${\bf
\bigvee}_{1} {\mathfrak W}({\mathfrak A}^{t_k})$ of the form $\widehat{\sigma}_{s}f^{s} -
\widehat{\sigma}_{t} \widehat{\eta}^{t}_{s} f^{s}$.  But for any such pair,
$\phi_{\mu}(\widehat{\sigma}_sf^s - \widehat{\sigma}_t \widehat{\eta}^t_s f^s) = \mu_s(f^s)
- \mu_t(\ehat f^s)) = (\mu_s - \eta^t_s \mu_t)(f^s) = 0.$ From the linearity of
$\phi_{\mu}$, any element of the linear span of such elements maps to zero, and therefore
any element of the closed span, since $\sum \mu_t(h^t) \leq \|h\|\hspace{3mm}\forall h =
(h^t)_{t \in {\mathbf J}} \in {\mathfrak M}.$

The functional $\phi_{\mu}$ is a state on \winf.  For note that
$\phi_{\mu}(\sigma_t(\chi_{\Omega})) = \mu_t(\chi_{\Omega}) =  1$. Moreover, $\phi_{\mu}
\geq 0,$ i.e., $[f] \geq 0$ implies that $\phi_{\mu}([f]) \geq 0$. In fact, from Lemma
III.3, $[0] \leq [f]$ iff for any $f \in [f]$, there exists $h \in {\mathfrak M}$ such
that $0 \leq f + h.$  Represent $f+h = \sum \widehat{\sigma}_{t}(f^{t} + h^{t})$.  By
definition of the $l_{1}$-join, $0 \leq f+h$ implies that $0 \leq f^{t} +
h^{t}\hspace{.5em} \forall t \in {\mathbf J}.$ Since $\mu_{t} \geq 0$ by definition,
$\forall t \in {\mathbf J}, \mu_{t}(f^{t} + h^{t}) \geq 0$, and hence $0 \leq \sum
\mu_t(f^t + h^t) = \phi_{\mu}([f]).$

Define $\beta = \alpha^{-1}.$ To show that $\beta$ is defined on all of $F_{\infty}$, fix
any $\phi \in F_{\infty}$, and define $\mu_t = \bigcirc$*$(\sigma_t)
\phi\hspace{3mm}\forall t \in {\mathbf J}$.  Then $\mu = (\mu_t)_{t \in {\mathbf J}}$ is
a thread, i.e., $\mu \in E_{\infty},$   from the composition rule in Proposition IV.4.
Note especially that $\phi(\sum^{\infty} \sigma_t(f^t)) = \sum^{\infty} \mu_t(f^t)$, from
the linearity and (strong) continuity of $\phi$.  For the composition rule for $\beta$,
one has by construction that $\phi_{\mu} \circ \sigma_{t} = \mu_{t}\hspace{.5em}\forall t
\in \mathbf {J}$. Then for any $t \in \mathbf {J}$,
$\bigcirc$*$(\sigma_{t})(\phi_{\mu})(f^{t}) = \phi_{\mu}(\sigma_{t}f^{t}) =
\mu_{t}(f^{t})\hspace{.5em}\forall f^{t} \in \wat{t},$ and therefore
$\bigcirc$*$(\sigma_{t})\phi_{\mu} = \mu_{t} = \rho_{t} \beta (\phi_{\mu}).$ This proves
the composition rule as stated.  To show that $\beta$ is injective, let $\phi\neq \psi$
be any two elements of  $F_{\infty}$.  Then there exists an $[f] \in {\mathfrak
W}^{\infty}$, say $[f] = \sum \sigma_{t}f^{t},$ such that $\phi([f]) \neq \psi([f])$.
Hence, for some $t \in {\mathbf J}$, $\bigcirc$*$(\sigma_{t})\phi \neq
\bigcirc$*$(\sigma_{t})\psi,$ so that $\rho_{t}(\beta \phi) \neq \rho_{t}(\beta \psi)$,
from the composition rule. Thus, $\beta \phi \neq \beta \psi.$

The proof is completed by showing that $\beta$ is continuous and therefore open, because
$F_{\infty}$ is compact. The subbasic neighborhoods (in the product topology) of $\mu \in
E_{\infty}$ are of the form $U(\mu; f^{t}, \varepsilon) = \{\nu \in E_{\infty} :
|\nu_{t}f^{t} - \mu_{t}f^{t}| < \varepsilon\}$ for any $t \in {\bf J}, f^{t} \in \wat{t},
\varepsilon > 0.$  The preimage of this set $\beta^{\leftarrow}(U(\mu; f^{t}, \varepsilon))
$ is $\{ \phi \in F_{\infty} : | \phi(\sigma_{t} f^{t}) - \phi_{\mu}(\sigma_{t}f^{t}| <
\varepsilon\}.$ But this is the standard form $N(\phi_{\mu}; \sigma_t f^t, \varepsilon)$ of
the subbasic neighborhoods of $\phi_{\mu}$ for the wk*-topology of $F_{\infty}.$ \qd

\begin{cor}
The set $F_{\infty}$ coincides with the compact set of states $\states$ on ${\mathfrak
W}^{\infty}.$
\end{cor}

\noindent \prf Note first that $\states$ is the intersection of a closed hyperplane and the
compact unit ball $\bigcirc$*$\winf$, and hence compact ({\em cf.} Proof, Proposition
II.1). We already have that $\beta^{-1} \mu = \phi_{\mu}$ is a state on \winf, and
therefore $F_{\infty} \subseteq \states$. For the reverse inclusion, fix any $\phi \in
{\mathcal K}({\mathfrak W}^{\infty})$.  Then $\bigcirc$*$(\sigma_{t}) \phi (\chi_{\Omega})
= \phi(\sigma_{t} \chi_{\Omega}) = \phi(e) = 1.$  For any $f^{t} \in \wat{t}$ with  $f^{t}
\geq 0$, $\sigma_{t}f^{t} = [f^{t}] \geq 0$, and hence $\bigcirc$*$(\sigma_{t}) \phi(f^{t})
= \phi([f^{t}]) \geq 0$ since $\phi \geq 0$.Therefore $\bigcirc$*$(\sigma_{t})\phi \in
E_{t}\hspace{.3em}\forall t \in \mathbf {J}$, and thus, $\phi \in
\bigcirc$*$(\sigma_{t})^{\leftarrow}(E_{t}) = F_{\infty}.$ Then $F_{\infty} \supseteq
{\mathcal K}({\mathfrak W}^{\infty}).$ \qd

Combination of Proposition IV.5 and its corollary yields $\states = F_{\infty} =
E_{\infty}$. We have thus proven the following.

\begin{thm} There is a 1:1 correspondence between the threads in $\threads$
and the elements of $\states$. It therefore makes sense to express this correspondence as
$\phi_{\mu} \leftrightarrow \mu$, subscripting the elements $\phi \in \states$ with the
corresponding threads $\mu \in \threads$. For any system $\Lambda_t$, the expectation value
of a measurement $f^t \in \wat{t}$ is fixed by the relation

\begin{equation}
\mu_t(f^t) = \phi_{\mu}(\sigma_tf^t)
\end{equation}

\spg i.e., by the value assigned to that property by the state $\phi_{\mu}$, for any state
$\mu \in E_{\infty}$.  The correspondence is defined by a unique homeomorphic bijection,
affine in both directions, between the compact spaces $K = \states$ and $E_{\infty}$.
\end{thm}

\spg {\em Proof.} To show that the transformation $\mu \mapsto \phi_{\mu}$ is affine, {\em
i.e.}, that for all $\mu,\nu \in \threads$ and for any $\lambda \in (0,1)$, $\lambda\mu +
(1-\lambda)\nu \mapsto \lambda\phi_{\mu}+(1-\lambda)\phi_{\nu}$, note that $\lambda\mu +
(1-\lambda)\nu \mapsto  \phi_{\lambda\mu + (1-\lambda)\nu}$, and for any finite set
$(t_k)^n$ of indices, $\phi_{\lambda\mu + (1-\lambda)\nu}(\sum^n \sigma_{t_k}(f^{t_k})) =
\sum^n(\lambda\mu_{t_k}+(1-\lambda)\nu_{t_k})(f^{t_k}) =
(\lambda\phi_{\mu}+(1-\lambda)\phi_{\nu})(\sum^n \sigma_{t_k}(f^{t_k}))$. For the other
direction, fix any $t \in {\mathbf J}$ and $f^t \in \wat{t}$. Then $(\lambda \phi_{\mu} +
(1-\lambda) \phi_{\nu})(\sigma_tf^t) = \lambda\phi_{\mu}(f^t) +
(1-\lambda)\phi_{\nu}(\sigma_tf^t) \rightarrow (\lambda\mu_t + (1-\lambda)\nu_t)(f^t) =
\rho_t(\lambda\mu-(1-\lambda)\nu)(f^t)$. \qd

\vspace{3mm} \begin{center} {\bf C. Identification of TL states with \threads.}
\end{center}

We conclude this section by showing the close relationship of the algebraic and TL
approaches announced in the Introduction. It is important to be able to apply theorems
from the algebraic structure such as the Choquet decomposability to the TL states.
However, this relationship is also needed by the algebraic theory itself. The calculation
of expectation values requires much more information about the lattice than is assumed by
the algebraic theory. Since this information is embodied in TL calculations, we may solve
the problem by applying the TL values themselves to the algebraic theory \emph{via} the
relationship of the programs.

By a TL state $\mu$ we mean an expectation-value operator for all bounded
Borel-measurable functions ${\mathfrak W}({\mathfrak A})$ on the phase space $\Omega$ of
the infinite lattice ({\em cf.} \cite[p.14]{ruel78}). One has the following
identification.

\begin{prop} Every TL state $\nu$ on $(\Omega, {\mathfrak A})$ is related to a unique
thread $\net \in E_{\infty}$ by its restrictions $\nu_t = \nu|_{\watt{t}}$ to the
individual $\wat{t}$, such that for any system $\Lambda_t$, $\nu_t(f^t) = \mu_t(f^t)$ for
all $f^t \in \wat{t}$.
\end{prop}

\noindent {\em Proof.} Consider the net of restrictions $(\nu_t)_{t \in {\bf J}}$, where
$\nu_t = \nu|_{\watt{t}} \forall t \in {\bf J}$. Homogeneity requires that for all $s, t
\in {\bf J}$ with $s \leq t$, the two equivalent observables $f^s \in \wat{s}$ and $\ehat
f^s \in \wat{t}$ have the same expectation value with respect to $\nu$, {\i.e.,} that
$\nu_s f^s = \nu_t\ehat f^s \hspace{3mm}\forall f^s \in \wat{s}$ and $\forall s \in {\bf
J}$ and $\forall t \geq s$. But this is eq.(4.2). Hence, $(\nu_t)_{t \in {\bf J}} \in
E_{\infty}$. The preceding proposition says that the net of projections is a unique
identification of the thread. \hspace{1cm} \qd

\noindent Combining this with the fact that the set of algebraic states is homeomorphic
with the set $\threads$, we now have  that each TL state is uniquely identifiable by its
expectation values with an algebraic state. We observe that the ability to form the
restrictions $\nu_t = \nu|_{\watt{t}}$ in this important result requires the construction
to be based on functions from the outside.

TL states are commonly described in terms of a transformation $\alpha_{\Lambda}:
\mathcal{K}{\mathfrak W}({\mathfrak A}) \rightarrow \mathcal{K}{\mathfrak W}({\mathfrak
A}_{\Lambda})$ that maps states on ${\mathfrak W}({\mathfrak A})$ to states on the set of
Borel-measurable functions on the configuration space $\Omega_{\Lambda}$ of any  finite
system $\Lambda$. To see that this is not something different, define
$\alpha^{\prime}_{\Lambda}: \Omega \rightarrow \Omega_{\Lambda}$ restricting $x \mapsto
x_{\Lambda}$ and $\alpha_{\Lambda}\sigma(f_{\Lambda}) = \sigma(f_{\Lambda} \circ
\alpha^{\prime})$ for all $f_{\Lambda} \in {\mathfrak W}({\mathfrak A}_{\Lambda})$ and
for any state $\sigma$ on ${\mathfrak W}({\mathfrak A})$ (\cite{ruel78}, p.14).

\begin{center} {\bf D. Indexing of algebraic states.}  \end{center}

\begin{prop}
The transformation $\delta_K: {\mathcal K}({\mathfrak W_K}) \rightarrow {\mathcal
K}{\mathcal C}(X_K)$ defined by $\delta_K x_{\mu}(f) = x_{\mu}(\psi_K^{-1}f)$ is an affine
homeomorphism onto $\mathcal{K}\mathcal{C}(X_K)$, and $\delta_K(X_K) =\partial_e{\mathcal
K}{\mathcal C}(X_K) $. Then $\delta_K$ extends the indexing of states by the definition
$\zeta_{\mu} = \delta_K x_{\mu} \hspace{2mm}\forall \mu \in \threads$.
\end{prop}

\noindent {\em Proof.}  The preceding proposition indexes ${\mathcal K}(\W_K)$ with
$x_{\mu}(\widehat{f}) \equiv \alpha_K \phi_{\mu} (\widehat{f}) = \widehat{f}(\phi_{\mu}).$
By Choquet's theorem, the relation $\widehat{f}(\phi_{\mu}) = \int_{\partial_eK}\widehat{f}
(\phi) d \sigma_{\mu}^{\prime} (\phi)$ uniquely identifies $\sigma_{\mu}^{\prime}$ with
$\phi_{\mu}$ and therefore $\mu$-indexes ${\mathcal S}(\partial_eK)$ in terms of the
homeomorphism ${\mathcal S}(\partial_eK)  = K$.  By the usual integral transformation
theorem \cite[Proposition 18.3.3]{sema71} and the fact that both $\alpha_K$ and $\psi_K$
are invertible, the mapping $\sigma_{\mu}^{\prime} \mapsto \sigma_{\mu}^{\prime}\circ
\alpha_K^{-1} \equiv \sigma_{\mu}$ is a bijection from ${\mathcal S}(\partial_eK)$ onto
${\mathcal S}(X_K)$ defining a $\mu$-indexing on ${\mathcal S}(X_K)$ as follows. One has,
for any $\mu \in \threads$ with representation in $K$,

\begin{equation}
\int_{X_K} f(x) d \sigma_{\mu}(x) = \int_{\partial_eK} f(\alpha_K
(\phi_{\nu})) d \sigma_{\mu}^{\prime}(\phi_{\nu})   \end{equation}

\noindent For all $\phi_{\nu} \in \partial_eK$,
$f(\alpha_K(\phi_{\nu}))  = \alpha_K \phi_{\nu} =
\widehat{f}(\phi_{\nu})$, and hence $f(x_{\nu}) =
x_{\nu}(\widehat{f}) = \widehat{f}(\phi_{\nu})$, where
$\widehat{f} = \psi_K^{-1}(f)$.  The integral becomes

\begin{equation}
\int_{X_K}  f(x) d \sigma_{\mu}(x) = \int_{\partial_eK}
\widehat{f}(\phi_{\nu}) d \sigma_{\mu}^{\prime} (\phi_{\nu}) =
\widehat{f}(\phi_{\mu})
\end{equation}

\noindent where the equality on the right is from Choquet's theorem again. That is,
$\sigma_{\mu} \in {\mathcal S}(X_K)$ is the unique probability measure on $X_K$ satisfying
this relation. By the Riesz representation theorem, $\mathcal{K}\mathcal{C}(X_K) =
{\mathcal S}(X_K)$, so that this result likewise $\mu$-indexes $\mathcal{K}\mathcal{C}(X_K)
$ with the definition $\zeta_{\mu}(f) = \widehat{f}(\phi_{\mu})$.

To define a mapping $\delta_K: {\mathcal K}(\W_K) \rightarrow \mathcal{K}\mathcal{C}(X_K)
$, note first that for all $x_{\mu} \in X_K$, $f(x_{\mu}) = x_{\mu}(\widehat{f}) =
\widehat{f}(\phi_{\mu})$. Hence, $\sigma_{\mu} = \delta(x_{\mu})$ (the Dirac point
functional), so that $\zeta_{\mu}(f) = f(x_{\mu}) \forall f \in \mathcal{C}(X_K)$.  This
is in fact a necessary and sufficient condition for $x_{\mu} \in X \equiv
\partial_e{\mathcal K}(\W)$.  Note that the condition defines a $\mu$-indexing
for the extremal states $X$. Define $\delta_K: X \rightarrow \mathcal{K}\mathcal{C}(X_K)$
by $\zeta_{\mu}(f) = \delta_K(x_{\mu})(f) = f(x_{\mu})$. Now for all convex combinations
$(a_n) \in \R$ and sets $(x_{\mu_n})_n \in X_K$, define $\delta_K(\sum a_n x_{\mu_n})  =
\sum a_n \delta_K(x_{\mu_n}).$ This extends $\delta_K$ to all of ${\mathcal K}(\W_K)$,
because by the Krein-Milman theorem, the compact convex set ${\mathcal K}(\W_K)$ is the
closed convex hull of its extremal points. Clearly, $\delta_K: {\mathcal K}(\W_K)
\rightarrow \mathcal{K }\mathcal{C}(X_K)$ is 1:1, because the $\mu$-indexing is unique.
Since it is affine, $\delta_K$ maps extremal points to extremal points. Since
$\mathcal{K}\mathcal{C}(X_K) = {\mathcal S}(X_K)$, and all Dirac point functionals
correspond to some $x_{\mu} \in X$, $\delta_K$ is onto $\mathcal{K}\mathcal{C}(X_K)$. To
show that it is also continuous and open, consider the wk*-subbasic set ${\mathcal
N}(x_{\mu}; \widehat{f}, \epsilon) = \{x_{\nu}: |x_{\nu}(\widehat{f}) -
x_{\mu}(\widehat{f})| < \epsilon \}.$  One has $\delta_K ({\mathcal N}(x_{\mu}:
\widehat{f}, \epsilon))  = \{\delta_K x_{\nu}: |\widehat{f}(\phi_{\nu}) -
\widehat{f}(\phi_{\mu})| < \epsilon \} = \{\delta_Kx_{\nu}: |\zeta_{\nu}(f) - \zemu(f)| <
\epsilon \} = {\mathcal N}(\zemu;f, \epsilon)$. That is, $\delta_K$ and $\delta_K^{-1}$
map subbasic sets onto subbasic sets. \qd

Eq. (4.5) allows us to write the exp.v. in  a familiar form. For any system $\Lambda_t$ and
observable $f^t \in \wat{t}$, let $\widehat{f}$ be the image of $f^t$ in $\W_K$, so that
for any $\mu_t \in E_t$, $\widehat{f}(\phi_{\mu}) = \mu_t(f^t)$. We then have

\begin{equation} \zemu (f) = \int_{X_K}f(x) d\sigma_{\mu}(x) = \mu_t(f^t)\hspace{3mm} \forall f
\in \Co ,  \end{equation}

\spg for all states $\zeta_{\mu} \in \Ck.$

In the algebraic QFT, the GNS construction defines a representation of the theory's
quasilocal observables as bounded linear operators on a certain abstract Hilbert space,
with expectation values calculated by inner products of the form $(\psi, A\psi)$.  That
is, the representation brings the algebraic theory into the form of ordinary quantum
mechanics.  We have now seen that the representation theorem in the classical algebraic
theory represents its quasilocal observables as $\Co$, continuous functions on a certain
compact ``phase space,'' with expectation values calculated as integrals over that space.
Thus, the  representation theorem brings the algebraic theory into the form of ordinary
CSM.

\vspace{5mm} \noindent{\bf {\large V \hspace{4mm} Applications }} \setcounter{section}{5}
\setcounter{equation}{0} \setcounter{thm}{0}

\vspace{3mm} Algebraic theory has to do with the abstract triple $\{\C,
\mathcal{K}\C,X\}$, where $X$ is a compact Hausdorff space. The role of the \hk  axioms
is to create a frame for interpreting mathematical conclusions about this triple in terms
of a particular underlying lattice problem. Let us display the whole hierarchy of spaces
defined in the algebraic construction:

\begin{center}
\begin{tabular}{c}

${\mathcal KC}(X_K) = {\mathfrak S}_K$\\

${\mathcal C}(X_K)$\\

$X_K \subset {\mathcal K}{\mathfrak W}_K  $\\

${\mathcal A}(K) \equiv {\mathfrak W}_K$\\

$K \subset \underline{\states = \threads} \overset{\rho_t}{\rightarrow} E_t$\\

$ \wat{t}   \overset{\sigma_t}{\rightarrow } \winf$\\

\end{tabular}
\end{center}

\spg We have underscored the equivalence of the threads and algebraic states in the
next-to-last line. (Recall in particular that the primary identification of the TL states
is with $E_{\infty}$.) The effect of the frame is to make everything above that line the
theory of a particular choice of \cc set $K \subset \states$. In this section and the next,
we study three distinct choices of $K$. The purpose is to  illustrate the importance of
this class of sets in physics and the effectiveness of the theory  in studying these sets
provided by the freedom in the choice of $K$.

\begin{center}   {\bf A. Compact convex sets}   \end{center}

The \cc sets arise in statistical mechanics because of their connection with  extremal
states. These  states are regarded as representing {\em pure thermodynamic phases} of a
problem. These states are readily identified in the algebraic setting as  the
multiplicative states on $\C$ \cite[Cor.4.5.4]{sema71}, the property that accounts for the
zero variance of observables in these states.   We may use the freedom in the choice of $K$
to match the algebraic problem with the physical problem as follows.

\begin{prop}  Fix any \cc set $K \subseteq \states$.  We may  define a set $X$ compact such
that the states on $\C$ are isomorphic with $K$, and $X$ to the set $\partial_eK$ of its
extremal states.  The triple $\{X , \C, {\mathcal K}\C\}$ so constructed is uniquely fixed
by either ${\mathcal K}\C$ or $X$.  For all states $\zeta \in {\mathcal K} (\C)$, there
exists a unique Radon probability measure $\sigma$ on $X$ such that

\[   \zeta (f) = \int_{X_K} f(x) d\sigma (x)\hspace{1cm}\forall f \in {\mathcal C}(X_K).  \]

\end{prop}

\spg {\em Proof.} Set ${\mathfrak W}_K = {\mathcal A}(K)$, and $X_K =\partial_e{\mathcal
K} ({\mathfrak W}_K)$. Then by Propositions III.15 and IV.9, $\delta_K \alpha_K(K) =
{\mathcal K}{\mathcal C}(X_K)$, and $X_K = \delta_K^{\leftarrow}(\partial_e{\mathcal
K}{\mathcal C}(X_K))$.   The set $\partial_e\Ck$ is identified as the set of
multiplicative states in $\Ck$.  By Proposition IV.9, the isomorphism $\delta^{-1}_X:\Ck
\rightarrow {\mathcal K}{\mathfrak W_K}$ maps $\partial_e\Ck$ onto $X_K$. Conversely, the
set of extremal states $\delta_K (X_K) =
\partial_e\Ck$ determines its closed convex hull $\Ck$ by the Krein-Milman Theorem. The
integral result is given by the Riesz Representation Theorem. \qd

The freedom  in matching the abstract algebra to particular problems afforded by this
Proposition is analogous to a flexibility in the QFT described by Emch \cite{emch72} as
the essential advantage of the algebraic approach over traditional theories based on Fock
space \cite[p.78]{emch72}. It is important to note that  the choice of $K$ in this
Proposition does not restrict the number of observables. In fact, Corollary III.13
assures that the algebra $\Co$ contains all the observables of the theory, for any $K$.
That is, each local observable $f^t \in \wat{t}$ maps to a unique element $f \in \Co$,
with its expectation value given by eq.(4.6).

Because of the identification of the extremal states with pure phases, the decomposition of
states into pure states is identified with phase separation. Clearly we expect  on physical
grounds that the most important states are the extremal states themselves or those states
that decompose into a small number of extremal states given by the Gibbs phase rule. Since
the extremal property must be defined with respect to a particular \cc set, the appearance
of extremal states signifies that the physical situation itself defines a certain \cc set
of states as {\em available} to the system, especially by the equilibrium condition. The
most common cases are spaces of states invariant under a particular symmetry, the
equilibrium (Gibbs) states, or an intersection of these.   According to the preceding
Proposition, if we set $K \subset \mathcal{K}\winf$ equal to the set of available states in
a particular problem, then {\em all} states on $\Co$ are ``available,'' and only these. We
illustrate these principles in the following applications.

\begin{center} {\bf B. Symmetry groups.}              \end{center}

The first application comes from the study of symmetries, following the   form and notation
of Ruelle \cite{ruel69}. A {\em symmetry} is an automorphism on the lattice that leaves the
expectation values of the theory unchanged. A {\em symmetry group} is a set of symmetries
with the group property. The symmetry groups are usually defined in terms of a group G, and
a transformation $\tau:G \rightarrow $aut$({\mathcal P})$ mapping  G to the automorphisms
on the set ${\mathcal P}$ of finite systems of the lattice. Since we are concerned with the
\cc sets of states $K \subset \states$, we need to transform automorphisms on the lattice
up to the set aut$(\winf)$ on $\winf$. Without danger of confusion, we use the same
notation $\tau_a$ to denote the corresponding transformation at each level. For simplicity,
we also fix, once and for all, a particular $a \in$ G.

The local transformations are as follows. Define $\tau_a: \wat{t} \rightarrow \wat{t}$ by
$\tau_a f^t(x) =  f^t(\tau^{-1}_ax) \hspace{1mm}\forall f^t \in \wat{t}$. For the states,
define $\tau_a: E_t \rightarrow E_t$ by $\tau_a \mu_t f^t = \mu_t(\tau_af^t)$. Now for
$\tau_a:$ G$ \rightarrow$aut($\winf$) itself,   the linear subspace $\mathfrak M$ is
generated by pairs of the form $\widehat{\sigma}_sf^s - \widehat{\sigma}_t \ehat f^s =
(\widehat{\sigma}_s - \widehat{\sigma}_t \ehat)f^s$. If we define $\tau_a\sigma_t =
\sigma_t\circ \tau_a \hspace{1mm}\forall t$, then $\tau_a(\widehat{\sigma}_sf^s -
\widehat{\sigma}_t \ehat f^s) =  (\widehat{\sigma}_s  - \widehat{\sigma}_t
\ehat)(\tau_af^s) \in \mathfrak M$. Hence $\tau_a {\mathfrak M} \subseteq {\mathfrak M}$,
{\em i.e.}, the subspace $\mathfrak M$ is closed under $\tau_a$. Then $\tau_a$ does not
disrupt equivalence classes, and we may define $\tau_a \in $ aut $ (\winf)$ by $\tau_a [f]
= [\tau_a f]$ on $\winf$.

Let ${\mathcal L}_G$ be the (closed) linear subspace of $\winf$ of elements of the form
$[g] = [f] - \tau_a[f]$ for any $a \in G$, and define the set of states ${\mathcal
L}_G^{\bot} = \{\phi \in \states: \phi[g] = 0, \hspace{2mm} [g] \in {\mathcal L}_G, a \in G
\}$. Clearly ${\mathcal L}_G^{\bot}$ is wk*-closed, $\phi$ linear, and therefore ${\mathcal
L}_G^{\bot}$ is a compact convex subset of $\states$. Then ${\mathcal L}_G^{\bot}$ is the
set of {\em G-invariant states}, {\em i.e.,} states with expectation values invariant under
transformations of the group $G$. Its extremal states $\partial_e({\mathcal L}_G^{\bot})$
are called the {\em G-ergodic states}. We take ${\mathcal L}_G^{\bot}$ as the set of
available states, and set $K = {\mathcal L}_G^{\bot}.$ Then the set of states on ${\mathcal
C}(X_K)$ is exactly the set of $G$-invariant states, and every G-invariant state admits a
unique decomposition into G-ergodic states $X_K = \partial_e\mathcal{K}\Co$.

The phenomenon of breakdown of symmetries gives a particularly clear picture   of available
states. For nested pairs of \cc sets $K_1 \subset K_2 \subset \mathcal{K}\winf$, the
extremal sets of $K_1$ are {\em not} generally extremal for $K_2$. Let the elements of
$K_1$  show a certain symmetry, and suppose the state $\phi \in K_1$ is extremal. Then
$\phi$ is a pure thermodynamic phase with that symmetry property if the only available
states are elements of $K_1$. But suppose instead that the set of available states is
$K_2,$ and $K_2$ does not possess this symmetry. We set $K = K_2$. If $\phi
\in\partial_eK_1 \cap(\partial_eK_2)^{\prime}$, then $\phi$ is no longer extremal, but
decomposes into elements of $\partial_eK_2$ that may not have the symmetry. We say that the
symmetry has been broken. The rule is as follows:  the opportunity for symmetry breakdown
arises whenever the invariant set is introduced into a larger set of available states that
are not all  invariant.

Now suppose the group G contains a subgroup H which is energetically favored, so that only
H-invariant states are available. We define as above ${\mathcal L}_G$ and ${\mathcal L}_H$.
Clearly, ${\mathcal L}_G \supset {\mathcal L}_H$. Since it is a stronger condition to be
invariant on the larger set, ${\mathcal L}_G^{\bot} \subset {\mathcal L}_H^{\bot}.$   Since
${\mathcal L}_H^{\bot}$ is now the availabe set, we take $K = {\mathcal L}_H^{\bot}$. Then
a G-ergodic state $\phi \in\partial_e{\mathcal L}_G^{\bot} \bigcap (\partial_e{\mathcal
L}_H^{\bot})^{\prime}$ will not be represented in $X_K$, {\em i.e.,} $\alpha_K\phi \not\in
X_K$. Hence, the state $\phi$ is not extremal, but is instead it   decomposed into
$H$-ergodic states in $X_K$. We say that the G-symmetry is broken.

\vspace{3mm} \begin{center} {\bf C. Gibbs states.}  \end{center}

The Gibbs states of the theory are identified as those threads $\mu \in \threads$ with
components $\mu_t \in E_t$ compatible with  assignment of a traditional Gibbs distribution
as a conditional distribution to each finite system in the space, as assured by the DLR
equations. One has the result from the TL program that a translation-invariant state is an
equilibrium state if, and only if, it is a Gibbs state \cite[Thm.4.2]{ruel78}. Denote the
invariant states on $\winf$ by $I$, and the set of all Gibbs states by $G$. Both are
compact convex sets. The invariant equilibrium states are the intersection $I \cap G$. With
$K = I \cap G$, the states in $\partial_e(I \cap G)$ are thermodynamic pure phases. But if
all Gibbs states are energetically available, then we set $K = G \subseteq \states$. Since
$I \cap G \subseteq G$, the above rule applies. Any invariant state in the intersection
$(\partial_e(I \cap G )) \cap (\partial_eG)^{\prime}$ decomposes into extremal Gibbs states
that are not invariant.   One says that the translational invariance of the theory is
broken \cite[4.3]{ruel78} .

\vspace{3mm} \begin{center} {\bf D. Statiomary states.}  \end{center}

\vspace{3mm} We conclude with the construction of the most basic set of states in classical
statistical mechanics, the stationary states. Let $E \subset \winf$ be the set of all
microcanonical (MC) states on the lattice, and let $K$ be the closed convex hull
$\overline{\mbox{co}}(E)$ of $E$. By MC states, we mean those states in $\winf$ identified
with TL states $\mu = \net \in \threads$ whose components are the projections of a given MC
state.

\begin{prop} The set $K$ is a \cc set of states, and $E = \partial_eK$.  \end{prop}

\noindent \prf The closed convex set $\overline{\mbox{co}}(E)$ is the same as the closure
of co$(E)$ \cite[Theorem V.2.4]{dunf57}.  But the closure of a convex set is convex
\cite[Theorem V..2.1]{dunf57}. Hence, $K$ is a \cc set, and we may use it to define the
triple $\{\Co, \mathcal{K}\Co, X_K\}$.  Now  $\phi_{\mu} \in K$ is an extremal state iff
$\zeta_{\mu} \in \mathcal{K}\Co$ is  extremal, for given $\mu \in \threads$. The extremal
states of $\mathcal{K}\Co$ are multiplicative, so that in particular, the energy density
has 0 variance on $X_K$. But this is true iff $\mu \in \threads$ is a MC state. \qd

The MC states are specified by pairs of values of the energy and particle-number densities,
related to the two constants of the motion. Since all stationary distributions are written
as Borel functions of these two constants, they may be regarded as distributions over the
set of MC states. Since $K$ is a \cc set, we may choose it to define $\Co$. Then by the
Riesz Representation Theorem, {\em the set $\mathcal{K}\Co$ consists of all distributions
on $X_K,$ and hence all stationary states on the lattice, including in particular the
traditional Gibbs equilibrium distributions}.

The set $X_K$ has the following remarkable structure.

\begin{thm}
The compact set $X_K \subset \mathcal{K}\mathfrak{W}_K$  is a finite set with the discrete
topology. All open sets $F \in \mathcal{B}(X_K)$ are clopen, and $X_K$ is extremely
disconnected.
\end{thm}

\noindent \prf  The set of transformations

\[
\wat{t} \stackrel{\sigma_t}{\rightarrow} \winf \stackrel{\Delta_K}{\rightarrow} \W
\stackrel{\psi_K} {\rightarrow} \Co
\]

\noindent represents local measurements on the system $\Lambda_t$ as the corresponding
quasilocal observables. Define $\gamma_t = \psi_K \circ \Delta_K \circ \sigma_t : \wat{t}
\rightarrow \Co$  for any $t \in {\mathbf J}$. Let $\g: X_K \rightarrow \R^2$ be defined by
$\g(x) = (H,N)$, where $H$ and $N$ are the energy and particle densities, respectively, of
the state $x$, and let $M = \g(X_K)$. Fix once and for all a finite system $\Lambda_t$. Let
$g^t \in \wat{t}$ and $n^t: \Omega \rightarrow \R$ be the energy and number densities,
respectively, for $\Lambda_t$, and define $\g^t: \Omega \rightarrow \R^2$ by $\g^t(x) =
(g^t(x), n^t(x)).$ Let $F \subset X_K$ be be any open set, and define $A_F = \g(F) \subset
M$.  Then the component $\mu_t \in E_t$ of thread $\net = \mu$ corresponding to a
particular state $x_{\mu} \in F$ is a MC  ensemble on the algebra of local observables
$\wat{t}$ corresponding to an energy and particle density in $A_F$. For any Borel set $B
\subset \R^2$, denote as usual $[\g^t \in B] = (\g^t)^{\leftarrow}(B)$, and Let
$\chi_{[\g^t \in B]}^{(\Omega)}:\Omega \rightarrow \{0,1\}$ be the characteristic function
of $[\g^t \in B]$ on the configuration space $\Omega$. Then $\chi_{[\g^t \in
A_F]}^{(\Omega)}(a) = 1$ if $\g^t(a) \in A_F$, and 0 otherwise. Clearly,
$\mu_t(\chi^{\Omega}_{[\g^t \in A_F]}) = 1$ if $x_{\mu} \in F$, and 0 otherwise, because
$[\g^t \in A_F]$ is the support of the component $\mu_t$ of any thread $\net = \mu$ for
which $x_{\mu} \in F$.  But $\gamma_t(f^t)(x_{\mu}) = \mu_t(f^t) \forall f^t \in \wat{t},
\mu \in \threads$, so that $\gamma_t(\chi_{[\g^t \in A_F]}^{\Omega})(x_{\mu}) = 1$ if
$x_{\mu} \in F$, and 0 otherwise.  Hence, $\gamma_t(\chi_{[\g^t \in A_F]}^{\Omega}) =
\chip{F}$. Thus, $\chip{F} \in \Co$.  But the  characteristic function $\chip{F}$ is
continuous iff $F$ is clopen.  Since $X_K$ is Hausdorff, the complement of any singleton
${x} \in X_K$ is open and therefore clopen.  Hence, all singletons are open, and $X_K$ is
discete. But the only discrete compact spaces are finite. \qd

\spg The compact extremely disconnected spaces are frequently called {\em Stonean}
spaces. Note especially that this theorem results from the algebraic structure itself,
without any assumptions about the topology of the lattice configuration space $\Omega$.

The Stonean topology for $X_K$ has the following consequence.   Let $\mathfrak{P} \subset
\Co$ be the lattice of idempotents in $\Co$. These are exactly the characteristic functions
of Borel sets in $X$, {\em i.e.,} functions of the form $\chip{B}(x) = 1, x \in B$, and 0
otherwise, where $B \subset X_K$ is a Borel set. The Stonean topology on $X_K$ is
equivalent to the condition that $\mathfrak{P}$ be a {\em complete} lattice \cite[Theorem
6.2d]{port88}.

\spg {\sc Acknowledgement}. The author wishes to express his gratitude to Rudolf Haag for
his many suggestions during the writing of this manuscript.

\end{document}